\documentclass[a4paper, 12pt]{article}
\pdfoutput=1

\newcommand{\blind}{1}

\usepackage{amsmath}
\usepackage{amssymb}
\usepackage{amsthm}
\usepackage{enumitem}
\usepackage{graphicx}
\usepackage{subcaption}
\usepackage{float}
\usepackage{natbib}
\setcitestyle{authoryear,round,semicolon,aysep={},yysep={,}}
\usepackage{xcolor}
\usepackage[linesnumbered, ruled,vlined]{algorithm2e}
\usepackage{multirow}
\usepackage{diagbox}
\usepackage{hyperref}
\hypersetup{
    colorlinks=true,
    linkcolor=blue,
    citecolor=blue,
    filecolor=magenta,      
    urlcolor=green,
}

\def\references{\bibliographystyle{rss}
\bibliography{sde}}

\newcommand{\Var}{\mathrm{Var}}
\newcommand{\Cov}{\mathrm{Cov}}
\newcommand{\argmin}{\mathop{\mathrm{argmin}}}

\def\T{{ \mathrm{\scriptscriptstyle T} }}
\newcommand{\mc}[1]{\mathcal{#1}}
\newcommand{\mb}[1]{\mathbb{#1}}

\newtheorem{thm}{Theorem}
\newtheorem{lem}{Lemma}
\newtheorem{crl}{Corollary}

\newcommand{\single}{\renewcommand{\baselinestretch}{1}\small\normalsize}
\newcommand{\double}{\renewcommand{\baselinestretch}{1.75}\small\normalsize}
\def\bco{\iffalse}

\def\hg #1 {\texcolor{cyan}{{\it Hans:}   #1}}

\begin{document}
\single


\if1\blind
{
  \title{\bf Dynamic Modeling of Sparse Longitudinal Data and Functional Snippets With Stochastic Differential Equations}
  \author{Yidong Zhou and Hans-Georg M\"{u}ller\thanks{
    This work was partially supported by NSF (DMS-2014626) and NIH 
    (ECHO).}\hspace{.2cm}\\
    Department of Statistics, University of California, Davis\\
    Davis, CA 95616, USA}
  \date{June 2024}
  \maketitle
} \fi

\if0\blind
{
  \bigskip
  \bigskip
  \bigskip
  \begin{center}
    {\LARGE\bf Dynamic Modeling of Sparse Longitudinal Data and Functional Snippets With Stochastic Differential Equations}
\end{center}
  \medskip
} \fi

\bigskip
\begin{abstract}
Sparse functional/longitudinal data have attracted widespread interest due to the prevalence of such data in social and life sciences. A prominent scenario where such data are routinely encountered are accelerated longitudinal studies, where subjects are enrolled in the study at a random time and are only tracked for a short amount of time relative to the domain of interest. The statistical analysis of such functional snippets is challenging since information for far-off-diagonal regions of the covariance structure is missing. Our main methodological contribution is to address this challenge by bypassing covariance estimation and instead modeling the underlying process as the solution of a data-adaptive stochastic differential equation. Taking advantage of the interface between Gaussian functional data and stochastic differential equations makes it possible to efficiently reconstruct the target process by estimating its dynamic distribution. The proposed approach allows one to consistently recover forward sample paths from functional snippets at the subject level. We establish the existence and uniqueness of the solution to the proposed data-driven stochastic differential equation and derive rates of convergence for the corresponding estimators. The finite sample performance is demonstrated with simulation studies and functional snippets arising from a growth study and spinal bone mineral density data.
\end{abstract}

\noindent%
\vfill

\double
\section{Introduction}
Functional data are commonly viewed as i.i.d. samples of realizations of an underlying smooth stochastic process, which  is typically  observed at a discrete grid of time points. Such data are common and routinely arise in longitudinal studies.  Functional data analysis has received much attention over the last decades, and  functional principal component analysis \citep{klef:73,cast:86,hall:06:1,chen:15:1,zhou:22} and functional regression \citep{rams:05,hall:07:1} have emerged as key tools. Detailed introductions and reviews can be found in \citet{rams:05,hsin:15,mull:16:3}. One area where there are still important open questions concerns the impact of the study design on the analysis. We develop here a novel type of analysis for functional snippets, which correspond to very sparse sampling designs that arise often in accelerated longitudinal studies, by establishing  a connection to stochastic differential equations (SDE).

From a general perspective, functional data are collected through various study designs, where one can differentiate between fully observed,  densely and sparsely sampled functional data  \citep{zhan:16}. Fully observed functional data occur in continuous sensor signal recordings and dense designs when measurements at a large number of well-spaced time points are available. Sparse designs are characterized by the availability of only a small number of measurements. A common sparse design occurs when sparse time points are distributed over the entire domain for each subject, 
with a smooth density that is strictly positive over the time domain where data are collected \citep{mull:05:4, mull:06:7,li:10}.

In this paper, the focus is on a  second type of sparse design that occurs in accelerated longitudinal studies \citep{galb:17}, where subjects are enrolled in the study at a random time within the time domain and are only tracked for a limited amount of time relative to the domain of interest. Such accelerated longitudinal designs are appealing for practitioners in social and life sciences as they minimize the time and resources required to collect data for each subject, especially when data gathering is costly, intrusive or difficult. Formally, denoting the domain of interest by $\mc{T}=[a, b]$, the $i$th subject is only observed on a sub-interval $[A_i, B_i]\subset\mc{T}$ where $B_i-A_i\leq\eta(b-a)$ for all $i$ and $\eta\in(0, 1)$ is a constant. Such data, when the constant $\eta$ is much smaller than 1, are referred to as functional snippets \citep{lin:21:2}.

Partially observed functional data also arise in the form of functional fragments \citep{krau:15, lieb:19, knei:20}, where the constant $\eta$ can be nearly as large as 1. The presence of large fragments makes such functional fragments easier to handle since the design plot \citep{mull:05:4} is typically fully or nearly fully covered by the design points, thereby enabling the estimation of the covariance surface directly from the data. In contrast, all of the design points for functional snippets fall within a narrow band around the diagonal area, where  the domain of interest is much larger than this band. It is therefore not possible to infer the covariance surface of the functional data with the usual nonparametric approach and this impedes the implementation of functional principal component analysis and all related methods.  The only known solution is to impose additional and typically very  strong and often unverifiable assumptions about the nature of the covariance. Such assumptions have been made to justify various forms of covariance completion that have included parametric, semiparametric and other approaches \citep{dela:16:3, desc:19, dela:21, lin:21:2, lin:22}.

For the modeling of time-dynamic systems, empirical dynamics for functional data \citep{mull:10:2} is an approach to recover the underlying dynamics from repeated observations of the trajectories that are generated by the dynamics, 
including a nonlinear version \citep{mull:12:6}. These approaches do not cover functional snippets. 
To the best of our knowledge, \citet{mull:18:6} is the only existing dynamic approach aimed at the analysis of  functional snippets, where the underlying dynamics are investigated through an autonomous differential equation for longitudinal quantile trajectories, requiring the underlying process to be monotonic \citep{mull:94:6, vitt:94} and is aimed at estimating the conditional quantile trajectories given an initial condition, rather than the whole dynamic distribution.

We aim to reconstruct the latent stochastic process that generates the observed functional snippets by recovering its time-evolving distributions, which we refer to as dynamic distribution. To overcome the challenge posed by snippets, we model the underlying process as the solution of a data-adaptive SDE. The dynamic distribution of the target process, containing all information about the underlying dynamics, is then estimated by stepwise forward integration. There is previous research \citep{comt:20, deni:21, pana:23} where functional data analysis has been utilized to infer SDEs, primarily focusing on the estimation of drift and diffusion coefficients with parametric components.  Our approach does not follow this approach and is entirely different, as our emphasis is the modeling of functional snippets and  on recovering the underlying stochastic process from such highly incomplete data. To accomplish this, we employ SDEs in a novel and fully nonparametric way.

The proposed SDE approach is novel and in contrast to various covariance completion approaches is nonparametric and does not involve functional principal component analysis. The latter requires to recover the complete covariance surface, which is straightforward for dense designs \citep{mull:00:1}, but for functional snippets in principle is impossible, unless one is prepared to impose strong assumptions about the global structure of the covariance that in general cannot be verified. The utility of the proposed approach for statistical practice is illustrated for growth and bone mineral density data in Section \ref{sec:data}, where it is shown to aid in growth monitoring and more generally distinguishing individuals with abnormal development patterns. The proposed approach includes  predictions for individuals with only one observation, where the individual-specific dynamic distribution far into the future can be obtained.  The rate of convergence for the corresponding conditional distributions is derived in terms of the Wasserstein metric. The main assumption of the proposed approach is that the underlying process is Gaussian, which is a common assumption in functional data analysis.  

The specific contributions of this paper are, first, to provide an alternative perspective to characterize functional snippets using SDEs;  
second, to recover future distributions of individual subjects under few assumptions; third, to provide an approach that works for minimal snippets, where only two adjacent measurements may be available for each subject; fourth, to demonstrate existence and uniqueness of the solution to the data-adaptive SDE,  along with the rate of convergence for the corresponding estimate; fifth, we demonstrate the wide applicability of the proposed dynamic modeling approach with growth snippets from Nepalese children and also with  bone mineral density data.

The rest of this paper is organized as follows. In Section \ref{sec:pre} we introduce the proposed dynamic model, while Section \ref{sec:est} covers estimation procedures. Theoretical results are established in Section \ref{sec:theory}. Simulations and applications for a Nepal growth study data and spinal bone mineral density data are discussed in Sections \ref{sec:fsp} and \ref{sec:data}, respectively. Finally, we conclude with a brief discussion in Section \ref{sec:dis}.

\section{Learning Dynamic Distribution via Stochastic Differential Equations}
\label{sec:pre}
\subsection{Stochastic differential equations and diffusion processes}
\label{sec:sto}
A typical (It\^{o}) stochastic differential equation (SDE) takes the form
\begin{equation}
	\label{eq:sde0}
	\begin{cases}
		dX_t=b(t, X_t)dt+\sigma(t, X_t)dB_t,\quad t\in\mathcal{T},\\X_0=x_0,
	\end{cases}
\end{equation}
where $X_t=X(t)$ is a stochastic process on $(\Omega, \mathcal{F}, P)$, $b$ and $\sigma$ are the drift and diffusion coefficients, respectively, and $B_t$ is a Brownian motion (also known as Wiener process). The initial value $x_0$ can be either deterministic or random, independent of the Brownian motion $B_t$. It is known that a unique solution of \eqref{eq:sde0} exists if the Lipschitz condition 
\begin{equation}
	\label{eq:lip}
	|b(t, x)-b(t, y)|+|\sigma(t, x)-\sigma(t, y)|\leq C|x-y|\quad\text{for all }x, y\in\mb{R}, t\in\mc{T}.
\end{equation}
and the linear growth condition
\begin{equation}
	\label{eq:growth}
	|b(t, x)|+|\sigma(t, x)|\leq C(1+|x|)\quad\text{for all }x\in\mb{R}, t\in\mc{T},
\end{equation}
hold for some constant $C>0$ \citep[chap. 5.2]{okse:13}. In fact, if coefficients $b$ and $\sigma$ satisfy the Lipschitz and linear growth conditions, then any solution $X_t$ is a diffusion process on $\mc{T}$ with drift coefficient $b$ and diffusion coefficient $\sigma$ \citep[p. 154]{pani:17}. A diffusion process is a continuous-time Markov process that has continuous sample paths, which can be defined by specifying its first two moments together with the requirement that there are no instantaneous jumps over time. We can write the formulae for the drift and diffusion coefficients of a diffusion process in the following form:
\begin{equation}
	\label{eq:drift}
	b(t, x)=\lim_{s\to t^+}\frac{1}{s-t}E(X_s-X_t|X_t=x)
\end{equation}
and 
\begin{equation*}
	\sigma^2(t, x)=\lim_{s\to t^+}\frac{1}{s-t}E\{(X_s-X_t)^2|X_t=x\}.
\end{equation*}
Note that the diffusion coefficient can be equivalently defined as 
\begin{equation}
	\label{eq:diffusion}
	\sigma^2(t, x)=\lim_{s\to t^+}\frac{1}{s-t}\Var(X_s-X_t|X_t=x)
\end{equation}
since 
\begin{align*}
	\Var(X_s-X_t|X_t=x)&=E\{(X_s-X_t)^2|X_t=x\}-\{b(t, x)(s-t)+o(s-t)\}^2\\&=E\{(X_s-X_t)^2|X_t=x\}+o(s-t).
\end{align*}
Here, $b(t, X_t)$ may be thought of as the instantaneous rate of change in the mean of the process given $X_t$; and $\sigma^2(t, X_t)$ can be viewed as the instantaneous rate of change of the squared fluctuations of the process given $X_t$ \citep[chap. 1.7]{kloe:99}. For a more detailed treatment, please refer to subsection S.1.1 of the Supplementary Material.

Diffusion processes originate in physics as mathematical models of the motions of individual molecules undergoing random collisions with other molecules  
\citep{pavl:14}. 
Brownian motion is the simplest and most pervasive diffusion process. Several more complex processes can be constructed from standard Brownian motion, including the Brownian bridge, geometric Brownian motion and the Ornstein-Uhlenbeck process \citep{uhle:30}. When drift and diffusion components  of a diffusion process are moderately smooth functions, its transition density satisfies partial differential equations, i.e., the Kolmogorov forward (Fokker-Planck) and the Kolmogorov backward equation. 

\subsection{Alternative formulation of stochastic differential equations}
\label{sec:alt}
We assume that the observed snippets are generated by an underlying stochastic process $X_t$ defined on some compact domain $\mathcal{T}\subset\mathbb{R}$ with mean function $\mu(t)=E(X_t)$, and covariance function $\Sigma(s, t)=\Cov(X_s, X_t)$. Without loss of generality, $\mathcal{T}$ is taken to be $[0, 1]$ in the sequel. Suppose $\{X_{t, 1}, \ldots, X_{t, n}\}$ is an independent random sample of $X_t$, where $n$ is the sample size. In practice, each $X_{t, i}$ is only recorded at subject-specific $N_i$ time points $T_{i1}, \ldots, T_{iN_i}$ and the observed data are $Y_{ij}=X_{T_{ij}, i}$ for $j=1, \ldots, N_i$. We assume that $N_i>1$ for the  subjects used to learn the SDE as subjects with only one measurement do not carry information about the local covariance structure. The snippet nature is reflected by the restriction that $|T_{ij}-T_{ik}|\leq\delta$ for all $i$, $j$, $k$, and some constant $\delta\in(0, 1)$. The focus of this paper is to infer stochastic dynamics of the underlying stochastic process $X_t$ from data pairs $(T_{ij}, Y_{ij})$,  $i=1, \ldots, n$,  $j=1, \ldots, N_i$. Specifically, we are interested in estimating sample paths of $X_t$ starting from some initial time point given a starting value. The proposed approach borrows information from subjects with at least two measurements in order to recover the  subject-specific dynamic distribution far into the future for each participant, even for those with a single measurement, which do not contribute to the model fitting step. To illustrate the effectiveness of the proposed method for snippets with minimal numbers of observations,  we consider the case $N_i= 2$ for simplicity. However, the proposed method is not restricted to this case and utilizes all data when more than two measurements are available for some or all subjects.  Additional details on this are in subsection \ref{subsec:nepal} and subsection S.5.3 of the Supplementary Material. 

The underlying stochastic process $X_t$ is assumed to follow a general SDE as per \eqref{eq:sde0}. In real data applications, the drift and diffusion coefficients in \eqref{eq:sde0} are typically unknown. To recover the underlying dynamics of $X_t$, instead of attempting to directly estimate the drift and diffusion terms, which is challenging for functional snippet data, we plug  in representations \eqref{eq:drift} and \eqref{eq:diffusion} of drift and diffusion coefficients to obtain the following alternative version of the SDE,
\begin{equation}
	\label{eq:sde}
	\begin{cases}
		dX_t=\frac{\partial}{\partial s}E(X_s|X_t)\Big|_{s=t}\cdot dt+\left\{\frac{\partial}{\partial s}\Var(X_s|X_t)\Big|_{s=t}\right\}^{1/2}\cdot dB_t,\quad t\in\mathcal{T},\\X_0=x_0.
	\end{cases}
\end{equation}
Note that $s$ is taken to be strictly greater than $t$ when calculating the partial derivatives of $E(X_s|X_t)$ and $\Var(X_s|X_t)$ with respect to $s$, in which case the diffusion coefficient is well-defined and not equal to 0.
SDE \eqref{eq:sde} is the key tool to obtain sample paths of $X_t$ given an initial condition 
by means of a recursive procedure, where under Gaussian assumption at each step the distribution of $X_t$ 
is constructed using the estimation of conditional means $E(X_s|X_t)$ and conditional variances $\Var(X_s|X_t)$.

Examples of the SDE as per \eqref{eq:sde} include Brownian motion, Ho-Lee model \citep{ho:86}, and Ornstein-Uhlenbeck process \citep{uhle:30}, among others. Different models postulate different forms of $b$ and $\sigma$. The Brownian motion $B_t$, with extensive applications in physics and electronics engineering, is a  special case of this SDE with zero drift and unit diffusion. The Ho-Lee model $dX_t=g(t)dt+\sigma dB_t$ with $\sigma>0$ and $g$ a deterministic function of time is a stochastic interest rate model widely used, for instance, for the pricing of bond options and to model  future interest rates. The Ornstein-Uhlenbeck process $dX_t=-\theta X_tdt+\sigma dB_t$ with $\theta>0, \sigma>0$ is often used to describe  mean-reverting phenomena in the physical sciences, evolutionary biology and finance. The coefficient $\theta$ characterizes the restoring force towards the mean and $\sigma$ characterizes the degree of volatility around the mean value.

\section{Estimation}
\label{sec:est}
\subsection{Simulating sample paths}
\label{em}
To estimate sample paths of $X_t$ given an initial condition from function snippets, it is instructive to rewrite the SDE in \eqref{eq:sde} as
\begin{align*}
&\lim_{s\to t^+}(X_s-X_t)\\&=\lim_{s\to t^+}\left\{\frac{E(X_s|X_t)-E(X_t|X_t)}{s-t}(s-t)+\left\{\frac{\Var(X_s|X_t)-\Var(X_t|X_t)}{s-t}\right\}^{1/2}(B_s-B_t)\right\}
\end{align*}
with an initial condition $X_0=x_0$. The above formula gives rise to a method to simulate the continuous-time process $X_t$ at a set of discrete time points given an initial condition. Consider a pre-specified equidistant time grid $0\leq t_0<t_1<\cdots<t_{K-1}<t_K\leq1$ with the common time spacing $\Delta$. Denote the initial value of $X_t$ at $t_0$ by $X_0$ and the simulation of $X_t$ at $t_k$ by $X_k$ for $k=1, \ldots, K$. We then simulate the continuous-time process $X_t$ at the discrete time points $t_k, k=1, \ldots, K$, given an initial condition $X_0=x_0$,  by the recursion
\begin{align*}
&X_k-X_{k-1}\\&=\frac{E(X_k|X_{k-1})-E(X_{k-1}|X_{k-1})}{\Delta}\Delta+\left\{\frac{\Var(X_k|X_{k-1})-\Var(X_{k-1}|X_{k-1})}{\Delta}\right\}^{1/2}(B_{t_k}-B_{t_{k-1}}).
\end{align*}
Observing that $E(X_{k-1}|X_{k-1})=X_{k-1}$, $\Var(X_{k-1}|X_{k-1})=0$, and $(B_{t_k}-B_{t_{k-1}})/\sqrt{\Delta}\sim N(0, 1)$, the above recursion reduces to
\begin{equation}
	X_k=E(X_k|X_{k-1})+\{\Var(X_k|X_{k-1})\}^{1/2}W_k,\quad X_0=x_0,
	\label{eq:em}
\end{equation}
where $W_k\sim N(0, 1)$ are independent for $k=1, \ldots, K$. 

We emphasize that under Gaussian assumption on the process $X_t$, the recursion in \eqref{eq:em} generates an exact simulation \citep{glas:04} of $X_t$ at $t_1, \ldots, t_K$ in the sense that the $X_k$ it produces follows the same distribution of the process $X_t$ at $t_k$ for all $k=1, \ldots, K$; see Lemma \ref{lem:exact} in Section \ref{sec:theory}. Classical simulation methods for SDEs, such as the Euler-Maruyama method and the Milstein method \citep{kloe:99}, in general, introduce discretization error at $t_1, \ldots, t_K$, because the increments do not have exactly the right mean and variance. To simulate $X_t$ using recursion \eqref{eq:em}, there is hence no need to consider increasing numbers of discrete time points $K$.  
In practice and particularly for the case of accelerated longitudinal studies, a good rule of thumb is to set the time spacing $\Delta$ as the scheduled (as opposed to actual) visit spacing for each subject. The number of discrete time points $K$ to simulate $X_t$ is then determined by the time spacing $\Delta$ and the time interval of interest; see Section \ref{sec:data} for the selection of time grids in real data applications.

To estimate sample paths of the process $X_t$, one needs to iteratively generate a random sample from $N\{E(X_k|X_{k-1}), \Var(X_k|X_{k-1})\}$ to simulate $X_t$ at $t_k$ for $k=1, \ldots, K$. In practice, both the conditional mean $E(X_k|X_{k-1})$ and conditional variance $\Var(X_k|X_{k-1})$ are unknown. One thus has to obtain their estimates to proceed. 

\subsection{Estimation of conditional mean and conditional variance}
\label{subsec:cmv}
Note that the information contained in $X_{k-1}=X_{t_{k-1}}$ is twofold and consists of  $X_{k-1}$ itself and also of the time index $t_{k-1}$. 
One can then formulate the estimation of the conditional mean $E(X_k|X_{k-1})$ and conditional variance $\Var(X_k|X_{k-1})$ as a regression problem with the response being $X_k$ and the predictor being $(X_{k-1}, t_{k-1})^\T$. 

Recall that each subject is observed at two time points $T_{i1}$ and $T_{i2}$, with measurements denoted by $Y_{i1}$ and $Y_{i2}$. Let $Z_i=(Y_{i1}, T_{i1})^\T$ and with a slight abuse of notation set $Y_i=Y_{i2}$ for $i=1, \ldots, n$. Viewing the  $\{(Z_i, Y_i)\}_{i=1}^n$ as $n$ i.i.d. realizations of the pair of random variables $(Z, Y)$, consider the regression model 
\begin{equation}
	\label{eq:re}
	Y_i=m(Z_i)+v(Z_i)\epsilon_i,
\end{equation}
where $m(z)=E(Y|Z=z)$ and $v^2(z)=\Var(Y|Z=z)$ are respectively the conditional mean function and conditional variance function. The error term $\epsilon_i$ satisfies $E(\epsilon_i|Z_i)=0$ and $\Var(\epsilon_i|Z_i)=1$. The estimation of both conditional mean and conditional variance using parametric or nonparametric regression models has been thoroughly studied. For the estimation of conditional variance we adopt the well-known approach of fitting a regression model for the squared residuals $\{Y_i-\hat{m}(Z_i)\}^2$ as responses and $Z_i$ as predictors \citep{fan:98:1}; see Section S.2 of the Supplementary Material for more details. 

Based on the regression model \eqref{eq:re}, the recursion for simulating sample paths in \eqref{eq:em} simplifies to
\begin{equation}
	X_k=m(Z_{k-1})+v(Z_{k-1})W_k,\quad X_0=x_0,
	\label{eq:em1}
\end{equation}
where $Z_{k-1}=(X_{k-1}, t_{k-1})^\T$ for $k=1, \ldots, K$. With estimates of the conditional mean function $\hat{m}(\cdot)$ and conditional variance function $\hat{v}^2(\cdot)$ in hand, we estimate the sample path of the underlying process $X_t$ at $t_1, \ldots, t_K$ given an initial condition $X_0=x_0$ using the following recursive procedure,
\begin{equation}
	\label{eq:simest}
	\begin{aligned}
		&\hat{X}_1=\hat{m}(Z_0)+\hat{v}(Z_0)W_1,\\&
		\hat{X}_k=\hat{m}(\hat{Z}_{k-1})+\hat{v}(\hat{Z}_{k-1})W_k,\quad k=2, \ldots, K,
	\end{aligned}
\end{equation}
where $Z_0=(x_0, t_0)^\T$ and $\hat{Z}_{k-1}=(\hat{X}_{k-1}, t_{k-1})^\T$ for $k=2, \ldots, K$; see Algorithm \ref{alg:em} for more details.

\begin{algorithm}[tb]
	\SetAlgoLined
	\KwIn{training data $\{(Z_i, Y_i)\}_{i=1}^n$, initial condition $Z_0=(x_0, t_0)^\T$, and time discretization $\{t_k, k=0, \ldots, K\}$.}
	\KwOut{$(\hat{X}_0, \ldots, \hat{X}_K)^\T$.}
	\For{$k=1, \ldots, K$}{
		Estimate the conditional mean $E(X_k|X_{k-1})$ and conditional variance $\Var(X_k|X_{k-1})$ by $\hat{m}(\hat{Z}_{k-1})$ and $\hat{v}^2(\hat{Z}_{k-1})$, respectively\;
		Draw a random sample $\hat{X}_k$ from $N\{\hat{m}(Z_{k-1}), \hat{v}^2(Z_{k-1})\}$\;
		$\hat{Z}_k\leftarrow(\hat{X}_k, t_k)^\T$\;
	}
	\caption{Estimating sample paths}
	\label{alg:em}
\end{algorithm}

If one has no prior knowledge about the conditional mean and conditional variance structure, which is often the case in real data applications, one will naturally adopt  nonparametric approaches that are more flexible than say multiple linear regression, while incurring a lower rate of convergence.  

\section{Theoretical Results}
\label{sec:theory}
We establish the existence and uniqueness of the solution to the proposed SDE as per \eqref{eq:sde} and the rate of consistency for the estimated sample path. The existence and uniqueness of the solution is built upon Gaussianity of the process $X_t$, i.e., for every finite set of time points $t_1, \ldots, t_k$ in $\mc{T}$, $(X_{t_1}, \ldots, X_{t_k})^\T$ are jointly Gaussian distributed. The key step is to express the conditional mean $E(X_s|X_t)$ and conditional variance $\Var(X_s|X_t)$ using the mean and covariance functions of $X_t$ whence drift and diffusion coefficients in \eqref{eq:sde} are seen to satisfy the Lipschitz and linear growth conditions as per \eqref{eq:lip} and \eqref{eq:growth}. Specifically, 
\begin{align}
	&E(X_s|X_t)=\mu(s)+\Sigma(s, t)\Sigma^{-1}(t, t)\{X_t-\mu(t)\},\label{eq:cm}\\
	&\Var(X_s|X_t)=\Sigma(s, s)-\Sigma(s, t)\Sigma^{-1}(t, t)\Sigma(t, s).\label{eq:cv}
\end{align}
If $X_t$ is non-Gaussian, as long as a unique solution exists, the rate of convergence for the estimated sample path can be similarly derived by assuming Lipschitz continuity for the conditional mean function $m(\cdot)$ and conditional variance function $v^2(\cdot)$; see Lemma \ref{lem:lip}.

To show that the drift and diffusion coefficients in \eqref{eq:sde} satisfy the Lipschitz and linear growth conditions as per \eqref{eq:lip} and \eqref{eq:growth}, we require the following conditions.
\begin{enumerate}[label=(A\arabic*)]
	\item The mean function $\mu(t)=E(X_t)$ is continuously differentiable on $\mathcal{T}$.\label{itm:mean1}
	\item The covariance function $\Sigma(s, t)=\mathrm{Cov}(X_s, X_t)$ is continuously differentiable in the lower triangular region $\{(s, t): s\geq t, s, t\in\mc{T}\}$. Equivalently, the two partial derivative functions of $\Sigma(s, t)$
	\[\Sigma_s'(s, t)=\frac{\partial\Sigma(s, t)}{\partial s},\;\Sigma_t'(s, t)=\frac{\partial\Sigma(s, t)}{\partial t}\]
	exist and are continuous for every $s, t\in\mc{T}$ and $s\geq t$.\label{itm:cov1}
\end{enumerate}
Conditions \ref{itm:mean1} and \ref{itm:cov1} are regularity conditions on the process $X_t$, where the latter implies that $\Sigma(s, t)$ is continuously differentiable in the upper triangular region $\{(s, t): s\leq t, s, t\in\mc{T}\}$, as well but  may not be differentiable across the diagonal $s=t$, which is for example a well-known property of Brownian motion.  It is easy to verify that all examples of  processes in subsection \ref{sec:alt} satisfy Conditions \ref{itm:mean1} and \ref{itm:cov1}; see Section S.1 of the Supplementary Material.
\begin{thm}
	\label{thm:sol}
	If the stochastic process $X_t$ is Gaussian, satisfies Conditions \ref{itm:mean1}, \ref{itm:cov1},  and the initial value $x_0$ is a random variable independent of the $\sigma$-algebra $\mc{F}_{\infty}$ generated by $\{B_s, s\geq0\}$ with $E(x_0^2)<\infty$,  then the stochastic differential equation \eqref{eq:sde} has a pathwise unique strong solution
	\[X_t=x_0+\int_0^t\frac{\partial}{\partial r}E(X_r|X_s)\Big|_{r=s}ds+\int_0^t\left\{\frac{\partial}{\partial r}\Var(X_r|X_s)\Big|_{r=s}\right\}^{1/2}dB_s,\quad t\in\mc{T}\]
	with the property that 
	\begin{equation}
		\label{eq:sp1}
		X_t\text{ is adapted to the filtration }\mc{F}_t^{x_0}\text{ generated by }x_0\text{ and }\{B_s, s\in[0, t]\}
	\end{equation}
	and
	\begin{equation}
		\label{eq:sp2}
		\sup_{t\in\mathcal{T}}E(X_t^2)<\infty.
	\end{equation}
\end{thm}
All proofs are given in Section S.3 of the Supplementary Material. The uniqueness of the solution is in the sense that if $X_t$ and $Y_t$ are two processes satisfying \eqref{eq:sde}, \eqref{eq:sp1}, and \eqref{eq:sp2} then
\[X_t=Y_t\quad\text{for all }t\in\mc{T}\quad\text{a.s.}\]
The solution $X_t$ found above is a strong solution because the version $B_t$ of Brownian motion is given in advance and the solution $X_t$ constructed from it is $\mathcal{F}_t^{x_0}$-adapted. The Gaussianity implies that $X_t$ must be governed by a narrow-sense linear SDE \citep{kloe:99},  where the drift coefficient $b(t, X_t)=a(t)X_t+c(t)$ and the diffusion coefficient is additive, i.e., $\sigma(t, X_t)=\sigma(t)$. Indeed, the drift and diffusion coefficients in $\eqref{eq:sde}$ under Gaussian assumption are
\begin{align*}
	&b(t, X_t)=\mu'(t)+\Sigma_s'(s, t)\big|_{s=t}\Sigma^{-1}(t, t)\{X_t-\mu(t)\},\\&
	\sigma(t, X_t)=\left\{\Sigma'(t, t)-2\Sigma_s'(s, t)\big|_{s=t}\right\}^{1/2},
\end{align*}
indicating the SDE \eqref{eq:sde} is narrow-sense linear. The general solution of a linear SDE can be found explicitly. Specifically, if $X_t$ is Gaussian, as a solution of $\eqref{eq:sde}$ it is of the form
\[X_t=\Phi(t)\left\{x_0+\int_0^tc(s)\Phi^{-1}(s)ds+\int_0^t\sigma(s)\Phi^{-1}(s)dB_s\right\}.\]
where $a(t)=\Sigma_s'(s, t)\big|_{s=t}\Sigma^{-1}(t, t), \,\, c(t)=\mu'(t)-\Sigma_s'(s, t)\big|_{s=t}\Sigma^{-1}(t, t)\mu(t)$ and $\Phi(t)=e^{\int_0^ta(s)ds}$.

An important feature of the recursion in \eqref{eq:em} is that it generates an exact simulation of $X_t$ at $t_1, \ldots, t_K$ \citep{glas:04} if $X_t$ is a Gaussian process. 
\begin{lem}
    \label{lem:exact}
    If the stochastic process $X_t$ is Gaussian, then the recursion in \eqref{eq:em} generates an exact simulation of the stochastic process $X_t$ at $t_1, \ldots, t_K$ in the sense that the distribution of the $X_1, \ldots, X_K$ it produces is precisely that of the continuous-time process $X_t$ at time points $t_1, \ldots, t_K$. 
\end{lem}
Lemma \ref{lem:exact} ensures that we do not have to worry about the discretization error while deriving the rate of convergence for the estimated sample path if $X_t$ is Gaussian. Regarding the asymptotic property of the estimated sample path as per \eqref{eq:simest}, we investigate the rate of convergence for $\hat{X}_K$, which also applies to $\hat{X}_k$ for any $k$. The proof relies on a recursive formula for the sequence $|\hat{X}_k-X_k|$, where the Lipschitz continuity of the conditional mean function $m(\cdot)$ and conditional variance function $v^2(\cdot)$ is utilized. To this end, we require the following conditions regarding the variance function $\Sigma(t, t)$ and the design of functional snippets.
\begin{enumerate}[label=(B\arabic*)]
    \item \label{itm:v}The variance function $\Sigma(t, t)$ is strictly positive on the half-open interval $(0, 1]$.
\end{enumerate}
Condition \ref{itm:v} is reasonable for real data applications, especially in our case where one is interested in modeling stochastic dynamics of the process $X_t$, and all example processes discussed in subsection \ref{sec:alt} satisfy Condition \ref{itm:v}; see Section S.1 of the Supplementary Material. Under Gaussianity, 
the conditional mean and conditional variance in recursion \eqref{eq:em1} reduces to 
\begin{align*}
	&m(Z_{k-1})=\mu(t_k)+\Sigma(t_k, t_{k-1})\Sigma^{-1}(t_{k-1}, t_{k-1})\{X_{k-1}-\mu(t_{k-1}))\},\\&
	v^2(Z_{k-1})=\Sigma(t_k, t_k)-\Sigma(t_k, t_{k-1})\Sigma^{-1}(t_{k-1}, t_{k-1})\Sigma(t_{k-1}, t_k),
\end{align*}
where $Z_{k-1}=(X_{k-1}, t_{k-1})^\T$ and $t_k=t_{k-1}+\Delta$ denotes the discrete time points used to simulate the sample path of the underlying process $X_t$.
\begin{lem}
    \label{lem:lip}
    If the stochastic process $X_t$ is Gaussian and satisfies \ref{itm:mean1}, \ref{itm:cov1}, and \ref{itm:v}, then for $k=2, \ldots, K$ the conditional mean and conditional variance in recursion \eqref{eq:em1} satisfy
    \begin{align*}
    	&|m(\hat{Z}_{k-1})-m(Z_{k-1})|\leq L|\hat{X}_{k-1}-X_{k-1}|,
    	\\&|v(\hat{Z}_{k-1})-v(Z_{k-1})|=0,
    \end{align*}
    where $L=\max_{t\in\{t_1, \ldots, t_{K-1}\}}|\Sigma(t+\Delta, t)\Sigma^{-1}(t, t)|$ and $Z_{k-1}=(X_{k-1}, t_{k-1})^\T, \hat{Z}_{k-1}=(\hat{X}_{k-1}, t_{k-1})^\T$.
\end{lem}
Lemma \ref{lem:lip} guarantees that the sequence $|\hat{X}_k-X_k|$ does not grow too fast, whence  one can bound $|\hat{X}_K-X_K|$ by recursion.  Lemma \ref{lem:lip} holds for all example processes discussed in subsection \ref{sec:alt} with Lipschitz constant $L=1$; see Section S.1 of the Supplementary Material for details.

To obtain the rate of convergence for the estimated sample path, one also needs to examine the asymptotic behavior of the conditional mean function estimate $\hat{m}(\cdot)$ and the conditional variance function estimate $\hat{v}^2(\cdot)$. Assume one has results  
for any fixed $z\in\mathbb{R}\times\mathcal{T}$ of the type 
\begin{equation}
	\label{eq:mvrate}
	[E\{|\hat{m}(z)-m(z)|^2\}]^{1/2}=O(\alpha_n),\quad [E\{|\hat{v}^2(z)-v^2(z)|^2\}]^{1/2}=O(\beta_n).
\end{equation}
If the residual-based estimator as described in Section S.2 of the Supplementary Material is adopted to estimate the conditional variance function $v^2(\cdot)$, it is well-known that the estimation of the conditional mean function $m(\cdot)$ has no influence on the estimation of $v^2(\cdot)$ \citep{fan:98:1}. Then $\beta_n=\alpha_n$ if the same regression method is performed to estimate $m(\cdot)$ and $v^2(\cdot)$. In the case of multiple linear regression, $\alpha_n=\beta_n=n^{-1/2}$, while $\alpha_n=\beta_n=n^{-1/3}$ for  local linear regression.

\begin{thm}
	\label{thm:strong}
	If the stochastic process $X_t$ is Gaussian and satisfies \ref{itm:mean1}, \ref{itm:cov1}, and \ref{itm:v}, then for the estimated sample path of the SDE \eqref{eq:sde} as defined in \eqref{eq:simest},
	\[\{E(|\hat{X}_K-X_K|^2)\}^{1/2}=O(\alpha_n+\beta_n),\]
	where $\alpha_n$ and $\beta_n$ are the rates of convergence for the conditional mean function estimate $\hat{m}(\cdot)$ and conditional variance function estimate $\hat{v}^2(\cdot)$ as per \eqref{eq:mvrate}.
\end{thm}
Theorem \ref{thm:strong} implies that $\hat{X}_K$ achieves strong convergence to $X_K$, where both the mean and variance converge, i.e.,
\[|E(\hat{X}_K)-E(X_K)|=O(\alpha_n+\beta_n),\quad|\Var(\hat{X}_K)-\Var(X_K)|=O(\alpha_n^2+\beta_n^2).\]
Note that this convergence holds uniformly over $k$, thereby establishing the pathwise convergence of the estimated sample path to the true process. 


Writing $\mathcal{L}(X_K)$, $\mathcal{L}(\hat{X}_K)$ for the distributions of $X_K$ and of the corresponding estimator $\hat{X}_K$, respectively, we aim to quantify the discrepancy between $\mathcal{L}(\hat{X}_K)$ and $\mathcal{L}(X_K)$ as a measure of the performance of the estimator. The strong convergence results obtained in Theorem \ref{thm:strong} can be used to obtain the rate of convergence of the $2$-Wasserstein distance \citep{vill:09}  
$d_W\{\mathcal{L}(\hat{X}_K),\mathcal{L}(X_K)\}$, where the 
$2$-Wasserstein distance between two probability measures $\nu_1$, $\nu_2$ on $\mathbb{R}$ is 
$d_W^2(\nu_1, \nu_2)=\int_0^1\{F_1^{-1}(p)-F_2^{-1}(p)\}^2dp,$
with  $F_1^{-1}$ and $F_2^{-1}$ denoting the quantile functions of $\nu_1, \nu_2$, respectively. If $\nu_1$ and $\nu_2$ are one-dimensional Gaussians with means and variances $(m_1, \sigma_1^2)$ and $(m_2, \sigma_2^2)$ then 
$d_W^2(\nu_1, \nu_2)=(m_1-m_2)^2+(\sigma_1-\sigma_2)^2.$ For the Wasserstein rate of convergence we obtain
\begin{crl}
	\label{thm:weak}
	Under the conditions of Theorem \ref{thm:strong}, the distribution of the estimated sample path as per \eqref{eq:simest} satisfies
	\[d_W\{\mathcal{L}(\hat{X}_K), \mathcal{L}(X_K)\}=O(\alpha_n+\beta_n),\]
	where $\alpha_n$ and $\beta_n$ are the rates of convergence for the conditional mean function estimate $\hat{m}(\cdot)$ and conditional variance function estimate $\hat{v}^2(\cdot)$ as per \eqref{eq:mvrate}.
\end{crl}

So far, we have assumed that snippets are observed without measurement errors, which applies to situations such as longitudinal growth curves, where anthropometric measurements are typically considered to be error-free. Applications to growth curves are highlighted in subsection~\ref{subsec:nepal} and subsection S.5.1 of the Supplementary Material using growth curve data for the Nepal and Berkeley growth studies. The presence of measurement errors will lead to an errors-in-variables scenario \citep{gril:86}, which will be discussed in Section S.4 of the Supplementary Material. There, we provide  theoretical analysis that characterizes the impact of measurement errors on the asymptotic behavior of the estimated sample paths. In subsection~\ref{subsec:sim}, we demonstrate that the proposed approach is quite robust to measurement errors. If one nevertheless would like to further address bias caused by  measurement errors, this will require adopting some of the available measurement error correction techniques \citep{cook:94,carr:06}. 

\section{Finite Sample Performance}
\label{sec:fsp}
\subsection{Implementation details}
\label{subsec:id}
The proposed dynamic modeling approach is straightforward to implement, as outlined in Algorithm~\ref{alg:em}. The regression model \eqref{eq:re} involves only a two-dimensional predictor, resulting in a time complexity of $O(n)$ for training. Consequently, Algorithm~\ref{alg:em} also runs in $O(n)$ time, given that calculating the conditional mean and conditional variance takes two steps and $K$ is fixed. For generating $M$ sample paths, the time complexity is $O(Mn)$, making it linear with respect to the sample size. This computational efficiency makes the proposed approach highly suitable for large datasets. The algorithm  has been implemented in \texttt{R} and is available on GitHub at \url{https://github.com/yidongzhou/Dynamic-Modeling-of-Functional-Snippets}.

Furthermore, the dynamic modeling approach inherently provides uncertainty quantification for the estimated sample paths. Practically, one can repeat the recursive process described in \eqref{eq:simest} $M$ times for a sufficiently large $M$, such as $M=1000$. With these $M$ simulated sample paths in hand,  an empirical $1-\alpha$ (pointwise) confidence band for the underlying process can be calculated. This method is demonstrated in subsection \ref{subsec:nepal} for identifying developmental delays in children's growth and is validated through simulations in subsection S.5.4 of the Supplementary Material. 

Note that nonparametric regression models, such as local linear regression, rely on two bandwidths, $h_1$ and $h_2$, for estimating the conditional mean and conditional variance, respectively, as defined in~\eqref{eq:re}. While $h_1$ can be selected via cross-validation, the bias in the squared residuals $\{Y_i-\hat{m}(Z_i)\}^2$ makes cross-validation infeasible for choosing $h_2$. In our implementation we choose $h_2=h_1$, where we select $h_1$ by cross-validation for conditional mean estimation, minimizing
\[\mathrm{CV}(h)=\sum_{l=1}^n\{Y_i-\hat{m}_h^{(-l)}(Z_l)\}^2.\]
Here $\hat{m}_h^{(-l)}(\cdot)$ denotes the local linear regression estimate using bandwidth $h$ based on the reduced sample $\{(Z_i, Y_i)\}_{i\neq l}$; users can choose to substitute alternative values for $h_1, h_2.$

\subsection{Simulation studies}
\label{subsec:sim}
We demonstrate the utility of the proposed approach in recovering underlying dynamics from functional snippets across various simulation contexts. Existing work based exclusively on covariance completion is not directly comparable, as one of the advantages of the proposed approach is that it bypasses covariance estimation and does not rely on functional principal component analysis. For comparative purposes, we estimate the covariance function using the covariance completion approach by \citet{lin:22}, denoted as LW, via the mcfda package available at \url{https://github.com/linulysses/mcfda}. Subsequently, assuming Gaussianity, we derive the conditional mean and conditional variance using the estimated mean and covariance functions. Finally, we apply the recursive procedure outlined in \eqref{eq:simest} to reconstruct the underlying stochastic process.

In our first example, we generate functional snippets from the Ho-Lee model and the Ornstein-Uhlenbeck process as discussed in subsection \ref{sec:alt}, respectively. To obtain functional snippets, we first simulate the sample path of $X_{t, i}$ at a regular time grid $\{t_k\}_{k=0}^K$ with $t_k=k\delta$ and $K\delta=1$ for each $i=1, \ldots, n$. Denoting the simulated values for the $n$ processes by $\{(X_{t_0, i}, X_{t_1, i}, \ldots, X_{t_K, i})^\T\}_{i=1}^n$, functional snippets are  generated as $\{(X_{T_i, i}, X_{T_i+\delta, i})^\T\}_{i=1}^n$ where for each $i$, $T_i$ is a time point randomly selected from the time grid $\{t_k\}_{k=0}^{K-1}$. Since both the Ho-Lee model and the Ornstein-Uhlenbeck process are narrow-sense linear SDEs, exact methods for simulating their paths are available by examining their explicit solutions \citep{glas:04}. Specifically, for the Ho-Lee model $dX_t=g(t)dt+\sigma dB_t$, a simple recursive procedure for simulating values at $\{t_k\}_{k=0}^K$ is
\begin{equation}
	\label{eq:hl}
	X_{k+1}=X_k+\int_{t_k}^{t_{k+1}}g(s)ds+\sigma(t_{k+1}-t_k)^{1/2}W_k,
\end{equation}
where the  $W_k\sim N(0, 1)$ are independent for all $k$ and $X_0=x_0$. Similarly for the Ornstein-Uhlenbeck process $dX_t=-\theta X_tdt+\sigma dB_t$, one can set
\begin{equation}
	\label{eq:ou}
	X_{k+1}=e^{-\theta(t_{k+1}-t_k)}X_k+\left\{\frac{\sigma^2}{2\theta}(1-e^{-2\theta(t_{k+1}-t_k)})\right\}^{1/2}W_k.
\end{equation}
The above procedures are exact in the sense that the joint distribution of the simulated values coincides with the joint distribution of the corresponding continuous-time process on the simulation grid. To investigate the effect of noise, we add independent errors to the generated functional snippets $\{(X_{T_i, i}, X_{T_i+\delta, i})^\T\}_{i=1}^n$. Specifically, we consider the contaminated functional snippets $\{(Y_{i1}, Y_{i2})^\T\}_{i=1}^n$ where $Y_{i1}=X_{T_i, i}+\varepsilon_{i1}$ and $Y_{i2}=X_{T_i+\delta, i}+\varepsilon_{i2}$ with $\varepsilon_{ij}\sim N(0, \nu^2)$ independently.

We examined the performance of the proposed approach across sample sizes $n=50, 200, 1000$ and noise levels $\nu=0, 0.01, 0.1$. For each combination of sample size and noise level, the simulation was repeated $Q=500$ times. The time interval was chosen as $[0, 1]$ and the time spacing was $\delta=0.05$. In each simulation, the recursive procedure as per \eqref{eq:simest} was performed $M=1000$ times using the contaminated functional snippets $\{(Y_{i1}, Y_{i2})^\T\}_{i=1}^n$ with the initial condition $Z_0=(0, 0)^\T$, from which $M=1000$ estimated sample paths evaluated at the time grid $\{t_k\}_{k=1}^K$ were obtained. We write $\{(\hat{X}_{t_1, l}, \ldots, \hat{X}_{t_K, l})^\T\}_{l=1}^M$ for the $M$ estimated sample paths. For each $l$, the corresponding true sample path $(X_{t_1, l}, \ldots, X_{t_K, l})^\T$ was obtained using the recursive procedure in \eqref{eq:hl} or \eqref{eq:ou} with the same initial value and $W_k$. For each run of a particular sample size and noise level, the quality of the estimation was quantified by the root-mean-square error,
\[\mathrm{RMSE}=\left\{\frac{1}{M}\sum_{l=1}^M(\hat{X}_{t_K, l}-X_{t_K, l})^2\right\}^{1/2}.\]

We chose $g(t)=\cos(t)$, $\theta=\sigma=1$ and the initial condition $X_0=0$ for the Ho-Lee model and the Ornstein-Uhlenbeck process, respectively, and used multiple linear regression to estimate the conditional mean and conditional variance for both cases. The mean and standard deviation of RMSE across $Q=500$ runs for various sample sizes and noise levels are summarized in Table~\ref{tab:armse}. We observe that the mean RMSE of the proposed approach diminishes as the sample size increases, while the presence of noise does not impact the results much. In contrast, the mean RMSE of the covariance completion method is substantial even with a sample size of 1000. This discrepancy may stem from the exceptionally sparse nature of this simulation scenario,  where each process is observed within a narrow window of length 0.05, contrasting sharply with the broader interval of interest, which is $[0, 1]$. Consequently, the available information may be too limited for covariance completion methods to accurately reconstruct the entire covariance surface. As shown in subsection S.5.3 of the Supplementary Material, the covariance completion approach performs better but is still inferior to the proposed approach when more measurements are available.

\begin{figure}[tb]
	\centering
	\includegraphics[width=\linewidth]{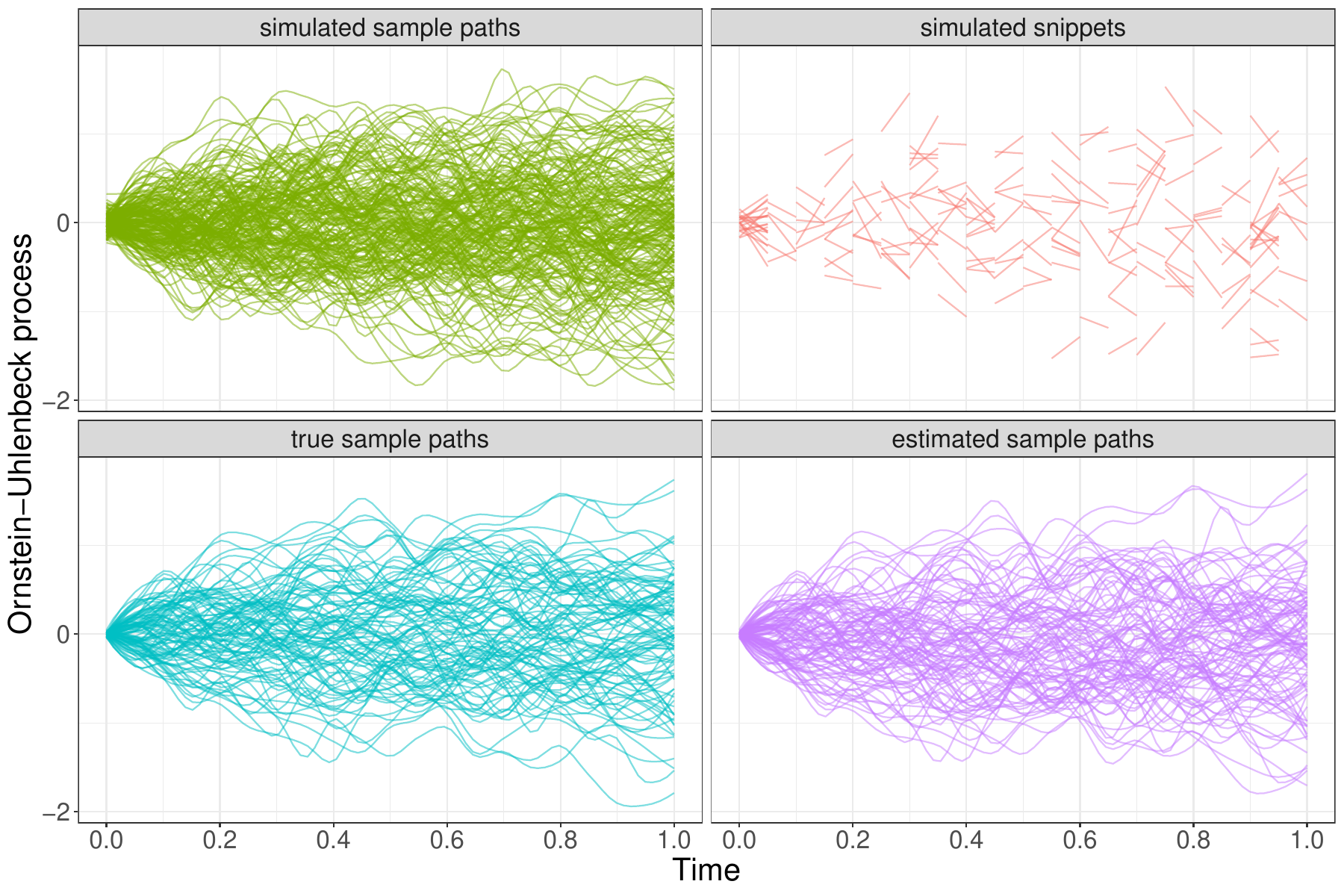}
	\caption{$M=100$ simulated sample paths (top left), simulated snippets (top right), true sample paths (bottom left), and estimated sample paths (bottom right) for the Ornstein-Uhlenbeck process. The sample size is $n=200$ and the noise level is $\nu=0.1$.}
	\label{fig:ou}
\end{figure}

\single
\begin{table}[tb]
	\footnotesize
	\centering
	\caption{Mean and standard deviation (in parentheses) of root-mean-square errors across 500 runs for the Ho-Lee model and the Ornstein-Uhlenbeck process. \textbf{DM}: the proposed dynamic modeling approach; \textbf{LW}: the covariance completion approach by \citet{lin:22}}
	\begin{tabular}{l|c|c|c|c|c|c}
		\hline
		\multirow{2}{*}{\diagbox[height=2\line]{\shortstack{Sample \\ size}}{\shortstack{Noise \\ level \\\vspace{-1em}}}} & \multicolumn{3}{c|}{DM} & \multicolumn{3}{c}{LW}\\\cline{2-7}
		& 0 & 0.01 & 0.1 & 0 & 0.01 & 0.1\\\hline
		\multicolumn{7}{l}{\textbf{Ho-Lee model}}\\\hline
		50 & 0.92 (0.87) & 0.93 (0.93) & 1.21 (1.71) & 1.08 (0.37) & 1.07 (0.34) & 1.11 (0.45)\\\hline
		200 & 0.39 (0.23) & 0.38 (0.22) & 0.54 (0.38) & 0.91 (0.34) & 0.92 (1.15) & 0.89 (0.27)\\\hline
		1000 & 0.17 (0.09) & 0.17 (0.09) & 0.27 (0.13) & 0.89 (0.24) & 0.89 (0.24) & 0.88 (0.25)\\\hline
		\multicolumn{7}{l}{\textbf{Ornstein-Uhlenbeck process}}\\\hline
		50 & 0.63 (0.65) & 0.62 (0.84) & 0.71 (1.81) & 0.72 (0.23) & 0.71 (0.2) & 0.74 (0.28)\\\hline
		200 & 0.25 (0.15) & 0.26 (0.16) & 0.27 (0.13) & 0.67 (0.17) & 0.66 (0.21) & 0.67 (0.23)\\\hline
		1000 & 0.11 (0.06) & 0.11 (0.06) & 0.15 (0.05) & 0.71 (0.09) & 0.70 (0.10) & 0.70 (0.11)\\\hline
	\end{tabular}
	\label{tab:armse}
\end{table}
\double
To further illustrate  the performance of the proposed dynamic modeling approach, we visualize the simulation results for the Ornstein-Uhlenbeck process with sample size $n=200$ and noise level $\nu=0.1$ in Figure~\ref{fig:ou}, where $M=100$ estimated sample paths are considered, along with the corresponding true sample paths. It is evident that the estimated sample paths  recover the underlying stochastic dynamics from very sparse  data, demonstrating that the proposed approach performs well.

\single
\double

Further simulations are provided  in Section S.5 of the Supplementary Material, where we emulate the Berkeley growth study data in the simulation set-up,  assess the resilience of the proposed method to departures from Gaussianity, and explore denser scenarios with $N_i=5$ measurements per subject. While the primary objective of the proposed approach is to reconstruct the dynamic distribution of the underlying process, we also investigate its capability for estimating the mean and variance function in subsection S.5.5 of the Supplementary Material.

\section{Data Applications}
\label{sec:data}
\subsection{Nepal growth study data}
\label{subsec:nepal}
Screening children's development status and 
monitoring height growth is essential for pediatric public health \citep{mull:12:4} and due to limited resources often must be based on incomplete data.  We demonstrate the potential of  the proposed dynamic modeling approach to characterize underlying growth patterns and reveal specific growth trends with snippet data from a  Nepal growth study  \citep{west:97}. This data set contains height measurements for 2258 children from rural Nepal taken at five adjacent times points  from birth to 76 months, spaced approximately four months apart. To facilitate the exploration of these data, we use the first 1000 records, containing measurements for 107 males and 93 females. Due to missing data,  the actual number of measurements per child ranges between 1 and 5. The children with at least two measurements in a row are included in the model, while the rest are used for model validation. We applied the proposed method to females and males separately since female and male growth trends differ significantly, with females reaching puberty far sooner than males.

Up to this point the number of measurements per subject $N_i$ was taken to be 2 for simplicity. 
For denser scenarios where $N_i>2$, one could divide the $N_i$ measurements into $N_i-1$ pairs of contiguous measurements for each $i$ and combine these pairs into a new sample for  conditional mean and conditional variance estimation. This is a useful approach to augment the sample size,  especially if the sample size $n$ is relatively small, which is often the case in practice. We employ this strategy to make full use of the Nepal growth study data as well as the spinal bone mineral density data  in the next subsection.

While in subsection S.5.1 of the Supplementary Material we demonstrated the efficacy of the proposed dynamic modeling approach to recover the  underlying growth dynamics from snippet data using Berkeley growth study data, we highlight here another important application of the proposed approach -- growth monitoring. Given a child's initial development status, the proposed approach predicts child-specific growth patterns far into the future. As a child grows older and fresh measurements become available, we are able to screen the child's development by comparing newly available measurements with the predicted growth. We demonstrate this with a  randomly selected female and  male who have no contiguous measurements and hence are not included in the model fitting. Specifically, the selected female was measured only once  at 4 months,  while the male was measured at 12 and  20 months. 

To obtain future growth patterns for these two children, the recursive procedure in \eqref{eq:simest} was implemented 100 times using the growth snippets with at least two measurements in a row to obtain 100 estimated growth curves, where local linear regression was  adopted to estimate the conditional mean and conditional variance. The starting time is $t_0=4$ months old for the selected female and $t_0=12$ months old for the selected male, where  the time spacing was set at $\Delta=4$ months, corresponding to the intended measurement spacing of the Nepal growth study. The starting height $X_0$ is chosen as the initial height measurement, i.e., 52.9 cm and 63 cm for the selected female and male, respectively. 

\single
\begin{figure}[tb]
	\centering
	\includegraphics[width=\linewidth]{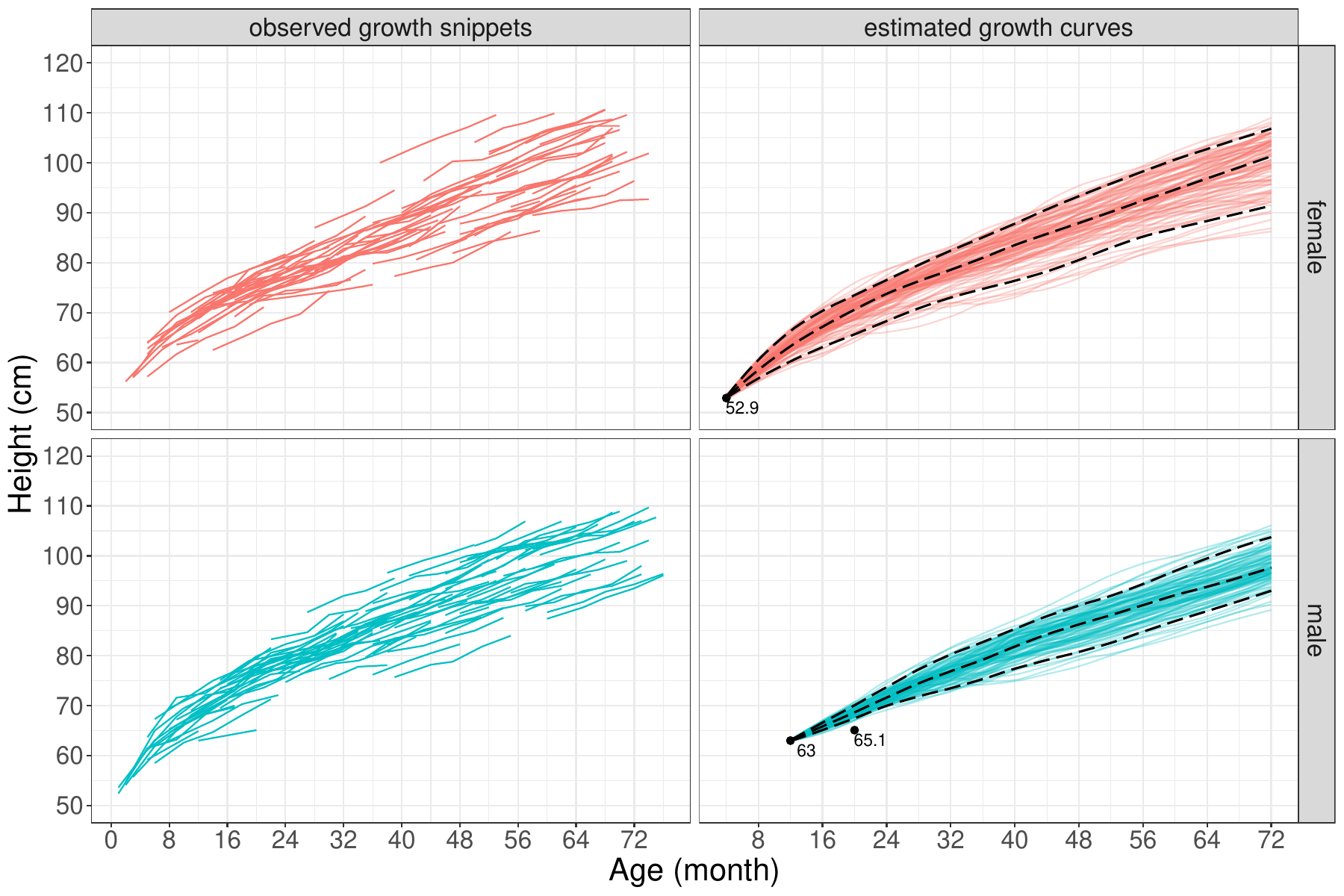}
	\caption{Observed growth snippets (left) and estimated growth curves (right) for the Nepal growth study data. The black dashed curves indicate $5\%, 50\%$ and  $95\%$ percentiles. Height measurements for the selected female and male are also highlighted.} 
	\label{fig:ngd}
\end{figure}
\double

The estimated growth curves and the corresponding $5\%, 50\%, 95\%$ percentile curves (dashed), along with the observed growth snippets are shown in Figure~\ref{fig:ngd}. Although the available information is very limited due to the sparse and snippet nature of the data, the proposed approach is capable of capturing relevant dynamics from the observed growth snippets and revealing future growth trends of the selected female and male. For the selected male, one fresh height measurement is available at later age (20 months old). In comparison to the estimated growth patterns, the newly available height measurement (65.1 cm) falls below $5\%$ percentile curve, indicating that this child may be developmentally delayed and require further follow-up.

\subsection{Spinal bone mineral density data}
In the study conducted by \citet{bach:99}, 423 healthy individuals underwent longitudinal assessments of their spinal bone mineral density. These assessments were scheduled annually over four consecutive years. However, deviations from the planned visit schedules resulted in irregularities in both the number of measurements per individual, ranging from 1 to 4, and the time intervals between measurements. The challenge posed by such irregular and sparse observations has garnered much attention in the field of functional data analysis \citep{jame:01, dela:16:3, dela:21, lin:21:2, lin:22}. We included 153 females and 127 males with ages ranging from 8.8 to 26.2 years with  at least 2 measurements  in model fitting, while the remaining subjects with only one measurement were used for model validation.

To infer individual-specific stochastic dynamics of spinal bone mineral density from the irregularly observed density snippets, we again randomly selected one female and male who were measured at ages 10 and 9 years, respectively.
Similarly, the recursive procedure in \eqref{eq:simest} was run 100 times to obtain 100 estimated bone density curves. Local linear regression was used to obtain conditional mean and variance, with cross-validation  bandwidth selection. The starting time was chosen as  $t_0=10$ years old for the selected female and $t_0=9$ years old for the selected male, with an end time of  $t_K=24$ years, with  one-year time increment, corresponding to the scheduled measurement spacing of the data. The starting values of bone mineral density are 0.778 and 0.642 for the selected female and male, respectively, corresponding to their initial density measurements.

\begin{figure}[tb]
	\centering
	\includegraphics[width=\linewidth]{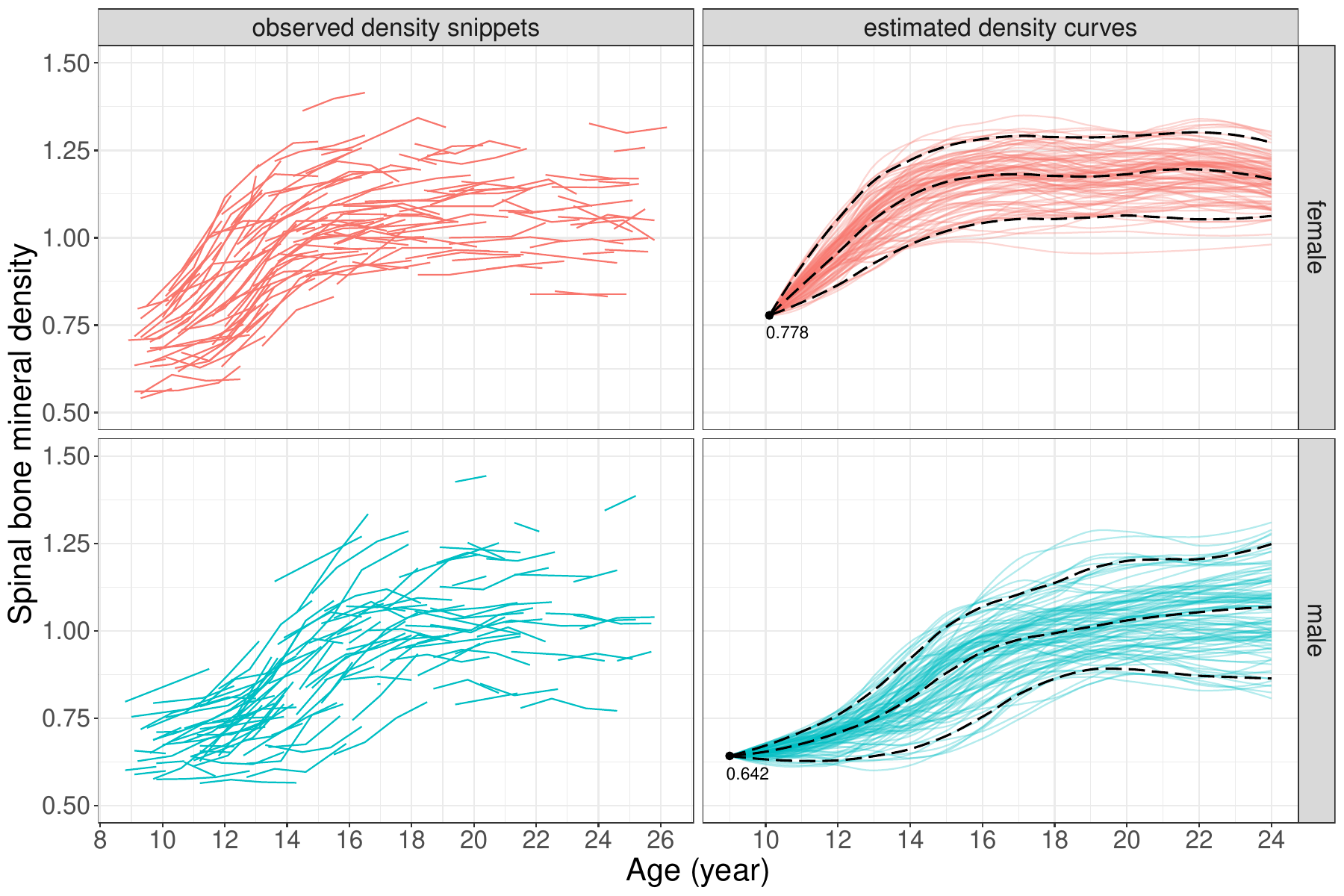}
	\caption{Observed growth snippets (left) and estimated bone density curves (right) for the spinal bone mineral density data. The black dashed curves indicate $5\%, 50\%$ and  $95\%$ percentiles. The bone density measurements of the selected female and male are also highlighted.}
	\label{fig:bmd}
\end{figure}

Figure~\ref{fig:bmd} depicts the observed snippets and estimated bone density curves, along with  $5\%, 50\%, 95\%$ percentile curves (dashed). 
We find that the proposed dynamic modeling approach accommodates the irregularity inherent in these data; see the right panel of Figure~\ref{fig:bmd}. 
Comparing the recovered bone density curves for the selected female and male, we find that for the female these curves reach a plateau at around 16 years,  while for the male they level off at around 18 years. This  finding  is in agreement with the literature \citep{bach:99}. For comparison purposes, we also applied the covariance completion approach to the spinal bone mineral density data and present the findings in Section S.6 of the Supplementary Material.

\single
\double

\section{Discussion}
\label{sec:dis}
In this article, we propose a flexible and robust approach to recover the dynamic distribution from functional snippets using stochastic differential equations (SDE). The proposed framework circumvents the challenge of estimating covariance surfaces in the presence of missing data in the off-diagonal regions, leading to a consistent reconstruction of sample paths from observed snippets. Both theoretical analysis and numerical simulations support the effectiveness and utility of the proposed SDE approach for analyzing functional snippets.

Differential equations are extensively used across various scientific fields, including engineering, physics, and biomedical sciences. A significant portion of the literature on differential equations focuses on parameter estimation \citep{lian:08}, with applications in time series \citep{chen:17} and functional data analysis \citep{deni:21}. Another research avenue involves neural differential equations, where differential equations enhance the performance of neural networks \citep{yada:15}. Notable examples include neural ordinary differential equations \citep{chen:18:2} and neural SDEs \citep{jia:19, oh:24}. Additionally, SDEs are applied in generative modeling, such as score-based diffusion models \citep{song:20}.

Complementing the existing literature, this paper uses SDEs as a powerful tool to model functional snippets and recover the underlying dynamics with minimal data on individual trajectories. The proposed tools also make it possible to assess forward dynamics by completing trajectories into the future when only minimal snippet information is available.

\single
\references
\double

\newpage


\begin{center}
{\large\bf SUPPLEMENTARY MATERIAL}
\end{center}
\setcounter{section}{0}
\renewcommand{\thesection}{S.\arabic{section}}
\renewcommand{\thesubsection}{S.\arabic{section}.\arabic{subsection}}
\section{Diffusion Processes and Examples}
\label{supp:exm}
\subsection{Drift and diffusion coefficients of diffusion processes}
Let $dt=s-t$ be an infinitesimal increment of time, and the corresponding increment of the stochastic process be $dX_t=X_s-X_t$. Using the SDE defined in \eqref{eq:sde0}, the infinitesimal mean is
\begin{align*}
	E(X_s-X_t|X_t=x)&=E\{b(t, x)(s-t)+\sigma(t, x)(B_s-B_t)\}+o(s-t)\\&=b(t, x)(s-t)+o(s-t).
\end{align*}
and the infinitesimal variance is
\begin{align*}
	E\{(X_s-X_t)^2|X_t=x\}&=E[\{b(t, x)(s-t)+\sigma(t, x)(B_s-B_t)\}^2]+o(s-t)\\&=\sigma^2(t, x)\Var(B_s-B_t)+E\{b(t, x)\sigma(t, x)(s-t)(B_s-B_t)\}\\&+E\{b^2(t, x)(s-t)^2\}+o(s-t)\\&=\sigma^2(t, x)(s-t)+o(s-t).
\end{align*}

For every positive real number $\varepsilon$, there exists a positive real number $\delta$ such that, for every $s>t$ such that $s-t<\delta$, we have
\[\left|b(t, x)-\frac{1}{s-t}E(X_s-X_t|X_t=x)\right|<\varepsilon.\]
Thus the limit
\[\lim_{s\to t^+}\frac{1}{s-t}E(X_s-X_t|X_t=x)\]
exists and equals $b(t, x)$. Similarly, we can show that the limit
\[\lim_{s\to t^+}\frac{1}{s-t}E\{(X_s-X_t)^2|X_t=x\}\]
exists and equals $\sigma^2(t, x)$ \citep[chap. 5]{pani:17}.

\subsection{Brownian motion}
The Brownian motion $B_t$ starting at $B_0=0$ is a Gaussian process with mean $E(B_t)=0$ and covariance $\Cov(B_s, B_t)=\min(s, t)$. It is straightforward to verify that $B_t$ satisfies Conditions \ref{itm:mean1}, \ref{itm:cov1} and \ref{itm:v}. Additionally, the Lipschitz constant in Lemma \ref{lem:lip} for Brownian motion is equal to $1$. 

\subsection{Ho-Lee model}
The Ho-Lee model defines the following SDE
\[dX_t=g(t)dt+\sigma dB_t,\]
where $\sigma>0$ and $g(t)$ is a deterministic function of time. Integrating both sides of the equation from $0$ to $t$ gives
\[X_t=X_0+\int_0^tg(s)ds+\sigma B_t.\]
Given $X_0=x_0$, it follows that
\[E(X_t)=x_0+\int_0^tg(s)ds\]
and 
\[\Cov(X_s, X_t)=\sigma^2\min(s, t).\]
Therefore, the mean function is continuously differentiable on $\mathcal{T}$ as long as $g(t)$ is continuous. In the lower triangular region $\{(s, t): s\geq t, s, t\in\mc{T}\}$, the covariance function is equal to $\sigma^2t$, which is also continuously differentiable. The variance function $\sigma^2t$ is positive on the half-open interval $(0, 1]$. We hence have shown that $X_t$ satisfies Conditions \ref{itm:mean1}, \ref{itm:cov1} and \ref{itm:v}. As for the Lipschitz constant in Lemma \ref{lem:lip}, one has
\[L=\max_{t\in\{t_1, \ldots, t_{K-1}\}}|\sigma^2t\cdot(\sigma^2t)^{-1}|=1.\]

\subsection{Ornstein-Uhlenbeck process}
The Ornstein-Uhlenbeck process is a solution of the  SDE
\[dX_t=-\theta X_tdt+\sigma dB_t,\]
with $\theta>0$, $\sigma>0$. Applying  It\^{o}'s formula to $e^{\theta t}X_t$, it is easy to show that
\[X_t=X_0e^{-\theta t}+\sigma\int_0^te^{-\theta(t-s)}dB_s.\]
Given $X_0=x_0$, one has
\[E(X_t)=x_0e^{-\theta t}\]
and 
\[\Cov(X_s, X_t)=\frac{\sigma^2}{2\theta}(e^{-\theta|s-t|}-e^{-\theta(s+t)}).\]
Hence, the mean function is continuously differentiable on $\mathcal{T}$. In the lower triangular region $\{(s, t): s\geq t, s, t\in\mc{T}\}$, the convariance function is equal to $\frac{\sigma^2}{2\theta}e^{-\theta s}(e^{\theta t}-e^{-\theta t})$, which is also continuously differentiable. The variance function $\frac{\sigma^2}{2\theta}(1-e^{-2\theta t})$ is positive on the half-open interval $(0, 1]$. We therefore have shown that $X_t$ satisfies Conditions \ref{itm:mean1}, \ref{itm:cov1} and \ref{itm:v}. Additionally, the Lipschitz constant in Lemma \ref{lem:lip} for the Ornstein-Uhlenbeck process is $L=e^{-\theta\delta}<1$.

\section{Conditional Variance Estimation}
\label{supp:cmcv}
Here we describe the residual-based estimator for the estimation of the conditional variance function $v^2(\cdot)$. Since $v^2(z)=E[\{Y-m(Z)\}^2|Z=z]$, we can regard the problem of estimating $v^2(\cdot)$ as a regression problem if the conditional mean function $m(\cdot)$ is given. In practice  $m(\cdot)$ is typically unknown and a  natural approach is to substitute  a regression estimator $\hat{m}(\cdot)$ for $m(\cdot)$.

Using standard notation of multiple linear regression, let $\mathbf{Z}$ be the $n\times2$ design matrix of the regression model in \eqref{eq:re} and $\mathbf{Y}$ the $n\times1$ response vector. If multiple linear regression is adopted to estimate $m(\cdot)$ and $v^2(\cdot)$, one has $\hat{m}(z)=z^\T\hat{\beta}$ for any $z=(x, t)^\T$, where $\hat{\beta}=(\mathbf{Z}^\T\mathbf{Z})^{-1}\mathbf{Z}^\T\mathbf{Y}$ is the least squares estimator. Now let $\hat{\mathbf{V}}$ be an $n\times1$ vector with the $i$th element being $\{Y_i-\hat{m}(Z_i)\}^2$. The residual-based estimator for $v^2(\cdot)$ is $\hat{v}^2(z)=z^\T\hat{\gamma}$ for any $z=(x, t)^\T$, with $\hat{\gamma}=(\hat{\mathbf{V}}^\T\hat{\mathbf{V}})^{-1}\hat{\mathbf{V}}^\T\mathbf{Y}$.

If instead local linear regression \citep{fan:96} is applied to estimate $m(\cdot)$ and $v^2(\cdot)$,  the conditional mean estimate is $\hat{m}(z)=\hat{\beta}_0$, where
\[(\hat{\beta}_0, \hat{\beta}_1)=\argmin_{\beta_0, \beta_1}\frac{1}{n}\sum_{i=1}^nK_{h_1}(Z_i-z)\{Y_i-\beta_0-\beta_1^\T(Z_i-z)\}^2.\]
Here $K_{h_1}(\cdot)=(h_{11}h_{12})^{-1}K(H_1^{-1}z)$ with $h_1=(h_{11}, h_{12})^\T$, $H_1=\mathrm{diag}(h_{11}, h_{12})$ and $K(\cdot)$ is a two-dimensional kernel corresponding to a symmetric two-dimensional probability density. Denote the squared residuals by $\hat{V}_i=\{Y_i-\hat{m}(Z_i)\}^2$. This leads to the residual-based estimator $\hat{v}^2(z)=\hat{\gamma}_0$ with bandwidth $h_2$, 
\[(\hat{\gamma}_0, \hat{\gamma}_1)=\argmin_{\gamma_0, \gamma_1}\frac{1}{n}\sum_{i=1}^nK_{h_2}(Z_i-z)\{\hat{V}_i-\gamma_0-\gamma_1^\T(Z_i-z)\}^2.\]
Bandwidths $h_1$ and $h_2$ can be selected data-adaptively using, for example, cross-validation.

For both multiple linear regression and local linear regression, the residual-based estimator $\hat{v}^2(\cdot)$ is not guaranteed to be nonnegative, and if it is negative we replace it by zero. 

\section{Proofs}
\label{supp:proof}

\subsection{Proof of Theorem \ref{thm:sol}}
\begin{proof}
	Let 
	\[b(t, X_t)=\frac{\partial}{\partial s}E(X_s|X_t)\Big|_{s=t},\quad\sigma(t, X_t)=\left\{\frac{\partial}{\partial s}\Var(X_s|X_t)\Big|_{s=t}\right\}^{1/2}.\]
	According to Theorem 5.2.1 in \citet{okse:13}, it suffices to show that the following linear growth and Lipschitz conditions are satisfied: for some constant $C>0$,
	\begin{align*}
		&|b(t, x)|+|\sigma(t, x)|\leq C(1+|x|)\quad\text{for all }x\in\mb{R}, t\in\mc{T},\\
		&|b(t, x)-b(t, y)|+|\sigma(t, x)-\sigma(t, y)|\leq C|x-y|\quad\text{for all }x, y\in\mb{R}, t\in\mc{T}.
	\end{align*}
	Observe that if $X_t$ is a Gaussian process, we have
	\begin{align*}
		&E(X_s|X_t)=\mu(s)+\Sigma(s, t)\Sigma^{-1}(t, t)\{X_t-\mu(t)\},\\
		&\Var(X_s|X_t)=\Sigma(s, s)-\Sigma(s, t)\Sigma^{-1}(t, t)\Sigma(t, s).
	\end{align*}
	Therefore
	\begin{align*}
		b(t, X_t)&=\left(\mu'(s)+\Sigma_s'(s, t)\Sigma^{-1}(t, t)\{X_t-\mu(t)\}\right)\big|_{s=t}\nonumber\\&=\mu'(t)+\Sigma_s'(s, t)\big|_{s=t}\Sigma^{-1}(t, t)\{X_t-\mu(t)\},\\
		\sigma(t, X_t)&=\left[\left\{\Sigma'(s, s)-2\Sigma_s'(s, t)\Sigma^{-1}(t, t)\Sigma(t, s)\right\}\big|_{s=t}\right]^{1/2}\nonumber\\&=\left\{\Sigma'(t, t)-2\Sigma_s'(s, t)\big|_{s=t}\right\}^{1/2}.
	\end{align*}
	Since $\mathcal{T}$ is compact, we have that $\mu(t)$, $\mu'(t)$, $\Sigma_s'(s, t)$, $\Sigma'(t, t)$ are bounded under Conditions \ref{itm:mean1} and \ref{itm:cov1}. Note also that if $\Sigma(t, t)=0$ then $\Sigma(s, t)=0$ for every $s\in\mc{T}$. Therefore, $\Sigma(t, t)=0$ indicates $\Sigma_s'(s, t)=0$ for every $s\in\mc{T}$ and thus $\Sigma_s'(s, t)\big|_{s=t}\Sigma^{-1}(t, t)$ is always bounded. For all $x\in\mb{R}$ and $t\in\mc{T}$, it follows that
	\begin{align*}
		|b(t, x)|+|\sigma(t, x)|&=\Big|\mu'(t)+\Sigma_s'(s, t)\big|_{s=t}\Sigma^{-1}(t, t)\{x-\mu(t)\}\Big|+\Big|\left\{\Sigma'(t, t)-2\Sigma_s'(s, t)\big|_{s=t}\right\}^{1/2}\Big|\nonumber\\&\leq C(1+|x|)
	\end{align*}
	for some constant $C$. The growth condition is hence satisfied. 
	For the Lipschitz condition, observe that $|\sigma(t, x)-\sigma(t, y)|=0$, we have
	\begin{align*}
		|b(t, x)-b(t, y)|+|\sigma(t, x)-\sigma(t, y)|&=|b(t, x)-b(t, y)|\\&=\Big|\Sigma_s'(s, t)\big|_{s=t}\Sigma^{-1}(t, t)(x-y)\Big|\\&=\Big|\Sigma_s'(s, t)\big|_{s=t}\Sigma^{-1}(t, t)\Big||x-y|\\&\leq C|x-y|
	\end{align*}
	for all $x, y\in\mb{R}$, $t\in\mc{T}$ and some constant $C$.
\end{proof}

\subsection{Proof of Lemma \ref{lem:exact}}
\begin{proof}
	Recall the recursion in \eqref{eq:em}. For any $k=1,\ldots, K$, one has
	\[X_k=E(X_k|X_{k-1})+\{\Var(X_k|X_{k-1})\}^{1/2}W_k.\]
	Under the assumption that $X_t$ is Gaussian, it suffices to show that the mean and variance of the right-hand side of the recursion are equal to $E(X_k)$ and $\Var(X_k)$, respectively. For the mean, 
	\begin{align*}
		E[E(X_k|X_{k-1})+\{\Var(X_k|X_{k-1})\}^{1/2}W_k]&=E(X_k)+E[\{\Var(X_k|X_{k-1})\}^{1/2}]\cdot E(W_k)\\&=E(X_k)
	\end{align*}
	since $W_k\sim N(0, 1)$ are independent. For the variance, note that $W_k$ is independent of $E(X_k|X_{k-1})$ and $\{\Var(X_k|X_{k-1})\}^{1/2}$. It follows that
	\begin{align*}
		\Var[\{\Var(X_k|X_{k-1})\}^{1/2}W_k]&=\{E(W_k)\}^2\Var[\{\Var(X_k|X_{k-1})\}^{1/2}]+\Var(W_k)E\{\Var(X_k|X_{k-1})\}\\&=E\{\Var(X_k|X_{k-1})\}
	\end{align*}
	and
	\begin{align*}
		\Cov[E(X_k|X_{k-1}),\{\Var(X_k|X_{k-1})\}^{1/2}W_k]&=E[E(X_k|X_{k-1})\cdot\{\Var(X_k|X_{k-1})\}^{1/2}W_k]\\&\hspace{1em}-E\{E(X_k|X_{k-1})\}\cdot E[\{\Var(X_k|X_{k-1})\}^{1/2}W_k]\\&=E[E(X_k|X_{k-1})\cdot\{\Var(X_k|X_{k-1})\}^{1/2}]\cdot E(W_k)\\&\hspace{1em}-E(X_k)\cdot E[\{\Var(X_k|X_{k-1})\}^{1/2}]\cdot E(W_k)\\&=0.
	\end{align*}
	Therefore,
	\begin{align*}
		\Var[E(X_k|X_{k-1})+\{\Var(X_k|X_{k-1})\}^{1/2}W_k]&=\Var\{E(X_k|X_{k-1}\}+\Var[\{\Var(X_k|X_{k-1})\}^{1/2}W_k]\\&\hspace{1em}+2\Cov[E(X_k|X_{k-1}),\{\Var(X_k|X_{k-1})\}^{1/2}W_k]\\&=\Var\{E(X_k|X_{k-1}\}+E\{\Var(X_k|X_{k-1})\}\\&=\Var(X_k).
	\end{align*}
\end{proof}

\subsection{Proof of Lemma \ref{lem:lip}}
\begin{proof}
Observe that under Gaussianity, the conditional mean and conditional variance in recursion \eqref{eq:em1} reduces to 
\begin{align*}
	&m(Z_{k-1})=\mu(t_k)+\Sigma(t_k, t_{k-1})\Sigma^{-1}(t_{k-1}, t_{k-1})\{X_{k-1}-\mu(t_{k-1}))\},\\&
	v^2(Z_{k-1})=\Sigma(t_k, t_k)-\Sigma(t_k, t_{k-1})\Sigma^{-1}(t_{k-1}, t_{k-1})\Sigma(t_{k-1}, t_k),
\end{align*}
where $Z_{k-1}=(X_{k-1}, t_{k-1})^\T$ and $t_k=t_{k-1}+\Delta$ denotes the discrete time points used to simulate the sample path of the underlying process $X_t$. Since $\mathcal{T}$ is compact, it follows that $\mu(t)$ and $\Sigma(s, t)$ are bounded under Conditions \ref{itm:mean1} and \ref{itm:cov1}. For $k=2, \ldots, K$, $\Sigma(t_k, t_{k-1})\Sigma^{-1}(t_{k-1}, t_{k-1})$ is bounded under Condition \ref{itm:v} since $t_{k-1}>0$ for $k=2, \ldots, K$ and $t_k=t_{k-1}+\Delta$. Recall that $\hat{Z}_{k-1}=(\hat{X}_{k-1}, t_{k-1})^\T$. For $k=2, \ldots, K$, we have
    \begin{align*}
        |m(\hat{Z}_{k-1})-m(Z_{k-1})|&=|\Sigma(t_k, t_{k-1})\Sigma^{-1}(t_{k-1}, t_{k-1})(\hat{X}_{k-1}-X_{k-1})|\\&\leq L|\hat{X}_{k-1}-X_{k-1}|,
    \end{align*}
    where $L=\max_{t\in\{t_1, \ldots, t_{K-1}\}}|\Sigma(t+\Delta, t)\Sigma^{-1}(t, t)|$. Since the conditional variance $v^2(Z_{k-1})$ is a function of $t_{k-1}$ only under Gaussian assumption, it immediately follows that $|v(\hat{Z}_{k-1})-v(Z_{k-1})|=0$.
\end{proof}

\subsection{Proof of Theorem \ref{thm:strong}}
\begin{proof}
Observe that the conditional variance 
\[\Var(X_s|X_t)=\Sigma(s, s)-\Sigma(s, t)\Sigma^{-1}(t, t)\Sigma(t, s),\quad 0\leq t<s\leq1\]
is nonnegative and equal to $0$ if and only if $X_s$ is deterministic or $X_s=cX_t$ for some constant $c\neq0$. The first case is excluded by Condition \ref{itm:v}, while the second case will not happen due to the presence of $dB_t$ in the SDE that $X_t$ follows. Therefore, $\Var(X_s|X_t)$ is always bounded away from zero for $0\leq t<s\leq1$ and hence the conditional variance function
\[v^2(z)=\Sigma(t+\delta, t+\delta)-\Sigma(t+\delta, t)\Sigma^{-1}(t, t)\Sigma(t, t+\delta)\]
is bounded away from zero for any fixed $z=(x, t)^\T$ and $t\in[0, 1-\delta]$.

It follows from \eqref{eq:mvrate} that $E\{|\hat{v}^2(z)-v^2(z)|^2\}=O(\beta_n^2)$. For any fixed $z=(x, t)^\T$ with $t\in[0, 1-\delta]$, one has
\begin{align}
    E\{|\hat{v}(z)-v(z)|^2\}&=E\left\{\frac{|\hat{v}^2(z)-v^2(z)|^2}{|\hat{v}(z)+v(z)|^2}\right\}\nonumber\\&\leq E\left\{\frac{|\hat{v}^2(z)-v^2(z)|^2}{v^2(z)}\right\}\nonumber\\&=\frac{1}{v^2(z)}E\{|\hat{v}^2(z)-v^2(z)|^2\}\nonumber\\&=O(\beta_n^2)\label{eq:vhatv}
\end{align}
since $\hat{v}(z)\geq0$ as per Section \ref{supp:cmcv} and $v^2(z)$ is bounded away from zero.

Recall the recursive procedure
\begin{align*}
    &\hat{X}_1=\hat{m}(Z_0)+\hat{v}(Z_0)W_1,\\&
    \hat{X}_k=\hat{m}(\hat{Z}_{k-1})+\hat{v}(\hat{Z}_{k-1})W_k,\quad k=2, \ldots, K,
\end{align*}
where $Z_0=(x_0, t_0)^\T$ and $\hat{Z}_{k-1}=(\hat{X}_{k-1}, t_{k-1})^\T$ for $k=2, \ldots, K$. For $k=1$, it follows that
\begin{align}
    E(|\hat{X}_1-X_1|^2)&=E[|\hat{m}(Z_0)-m(Z_0)+\{\hat{v}(Z_0)-v(Z_0)\}W_1|^2]\nonumber\\&=E\{|\hat{m}(Z_0)-m(Z_0)|^2\}+E[|\{\hat{v}(Z_0)-v(Z_0)\}W_1|^2]+\nonumber\\&\hspace{1.3em}2E(\{\hat{m}(Z_0)-m(Z_0)\}[\{\hat{v}(Z_0)-v(Z_0)\}W_1])\nonumber\\&=E\{|\hat{m}(Z_0)-m(Z_0)|^2\}+E\{|\hat{v}(Z_0)-v(Z_0)|^2\}E(W_1^2)\nonumber\\&=E\{|\hat{m}(Z_0)-m(Z_0)|^2\}+E\{|\hat{v}(Z_0)-v(Z_0)|^2\}\nonumber\\&=O(\alpha_n^2+\beta_n^2)\label{eq:x1}
\end{align}
since $W_1\sim N(0, 1)$ are independent of $\hat{m}(Z_0)$ and $\hat{v}(Z_0)$. 

Now, for any $k=2, \ldots, K$, one can similarly show that
\begin{align}
    E(|\hat{X}_k-X_k|^2)&=E[|\hat{m}(\hat{Z}_{k-1})-m(Z_{k-1})+\{\hat{v}(\hat{Z}_{k-1})-v(Z_{k-1})\}W_k|^2]\nonumber\\&=E\{|\hat{m}(\hat{Z}_{k-1})-m(Z_{k-1})|^2\}+E\{|\hat{v}(\hat{Z}_{k-1})-v(Z_{k-1})|^2\}\label{eq:xk}
\end{align}
where $Z_{k-1}=(X_{k-1}, t_{k-1})^\T$ since $W_k\sim N(0, 1)$ are independent of $\hat{m}(\hat{Z}_{k-1})$ and $\hat{v}(\hat{Z}_{k-1})$. Observe that
\[\hat{m}(\hat{Z}_{k-1})-m(Z_{k-1})=\{\hat{m}(\hat{Z}_{k-1})-m(\hat{Z}_{k-1})\}+\{m(\hat{Z}_{k-1})-m(Z_{k-1})\},\]
where the first term satisfies
\[E\{|\hat{m}(\hat{Z}_{k-1})-m(\hat{Z}_{k-1})|^2\}=O(\alpha_n^2)\]
by \eqref{eq:mvrate}, and the second term satisfies
\[|m(\hat{Z}_{k-1})-m(Z_{k-1})|\leq L|\hat{X}_{k-1}-X_{k-1}|\]
by Lemma \ref{lem:lip}. It follows that
\[E\{|m(\hat{Z}_{k-1})-m(Z_{k-1})|^2\}\leq L^2E(|\hat{X}_{k-1}-X_{k-1}|^2).\]
Therefore,
\begin{align*}
	E\{|\hat{m}(\hat{Z}_{k-1})-m(Z_{k-1})|^2\}&=E[|\{\hat{m}(\hat{Z}_{k-1})-m(\hat{Z}_{k-1})\}+\{m(\hat{Z}_{k-1})-m(Z_{k-1})\}|^2]\\&=O\{\alpha_n^2+L^2E(|\hat{X}_{k-1}-X_{k-1}|^2)\}.
\end{align*}

For the second term in \eqref{eq:xk}, one has
\begin{align*}
	\hat{v}(\hat{Z}_{k-1})-v(Z_{k-1})&=\hat{v}(\hat{Z}_{k-1})-v(\hat{Z}_{k-1})+v(\hat{Z}_{k-1})-v(Z_{k-1})\\&=\hat{v}(\hat{Z}_{k-1})-v(\hat{Z}_{k-1})
\end{align*}
since $v(\hat{Z}_{k-1})-v(Z_{k-1})=0$ by Lemma \ref{lem:lip}. It follows from \eqref{eq:vhatv} that
\begin{align*}
	E\{|\hat{v}(\hat{Z}_{k-1})-v(Z_{k-1})|^2\}&=E\{|\hat{v}(\hat{Z}_{k-1})-v(\hat{Z}_{k-1})|^2\}\\&=O(\beta_n^2).
\end{align*}
Therefore, 
\[E(|\hat{X}_k-X_k|^2)=O\{L^2E(|\hat{X}_{k-1}-X_{k-1}|^2)+\alpha_n^2+\beta_n^2\}.\]
Using the above recursion and \eqref{eq:x1} one has
\[E(|\hat{X}_k-X_k|^2)=O(\alpha_n^2+\beta_n^2)\]
for any $k=2, \ldots, K$. Hence,
\[\{E(|\hat{X}_K-X_K|^2)\}^{1/2}=O(\alpha_n+\beta_n).\]
\end{proof}

\subsection{Proof of Corollary \ref{thm:weak}}
\begin{proof}
	According to Theorem \ref{thm:strong}, $\{E(|\hat{X}_K-X_K|^2)\}^{1/2}=O(\alpha_n+\beta_n)$. It follows that
	\begin{equation*}
		E(|\hat{X}_K-X_K|)=\{\Var(|\hat{X}_K-X_K|)\}^{1/2}=O(\alpha_n+\beta_n).
	\end{equation*}
	If the stochastic process $X_t$ is Gaussian, one has
	\begin{align*}
		d_W^2\{\mathcal{L}(\hat{X}_K), \mathcal{L}(X_K)\}&=|E(\hat{X}_K)-E(X_K)|^2+|\{\Var(\hat{X}_K)\}^{1/2}-\{\Var(X_K)\}^{1/2}|^2\\&=R_1^2+R_2^2,
	\end{align*}
	where $R_1=|E(\hat{X}_K)-E(X_K)|$ and $R_2=|\{\Var(\hat{X}_K)\}^{1/2}-\{\Var(X_K)\}^{1/2}|$. For $R_1$ one has
	\begin{align*}
		R_1&=|E(\hat{X}_K)-E(X_K)|\\&\leq E(|\hat{X}_K-X_K|)\\&=O(\alpha_n+\beta_n).
	\end{align*}
	For $R_2$, it holds that
	\begin{align*}
		R_2&=|\{\Var(\hat{X}_K)\}^{1/2}-\{\Var(X_K)\}^{1/2}|\\&\leq|\Var(\hat{X}_K)-\Var(X_K)|^{1/2}\\&\leq\{\Var(|\hat{X}_K-X_K|)\}^{1/2}\\&=O(\alpha_n+\beta_n).
	\end{align*}
	Therefore,
	\begin{align*}
		d_W\{\mathcal{L}(\hat{X}_K), \mathcal{L}(X_K)\}&=R_1+R_2\\&=O(\alpha_n+\beta_n).
	\end{align*}
\end{proof}

\section{Impact of Measurement Errors}
\label{supp:mr}
Consider measurements with errors 
\begin{equation}
	\label{eq:err}
	\hat{Y}_{ij}=Y_{ij}+e_{ij},\quad i=1, \ldots, n, j=1, \ldots, N_i,
\end{equation}	
where $Y_{ij}=X_{T_{ij}, i}$ and $E(e_{ij})=0$. Instead of imposing a specific distributional assumption on the errors, such as the normal distribution, we adopt the small error variance assumption commonly used in deconvolution problems \citep{hall:02:1, carr:04}. Specifically, the variance of the errors is considered to tend to zero as the sample size increases, i.e., $\Var(e_{ij})\to0$ as $n\to\infty$. Asymptotics based on this small error variance assumption can effectively reflect finite sample scenarios where the error variance is relatively small compared to the variance of $Y_{ij}$ \citep{dela:16:1}.

\begin{thm}
	\label{thm:error}
	Suppose the stochastic process $X_t$ is Gaussian, \ref{itm:mean1}, \ref{itm:cov1}, \ref{itm:v} hold, and the measurement error $e_{ij}$ as per \eqref{eq:err} satisfies $\Var(e_{ij})=\zeta_n^2\to0$ as $n\to\infty$. For the estimated sample path of the SDE \eqref{eq:sde} as defined in \eqref{eq:simest}, it holds that
	\[\{E(|\hat{X}_K-X_K|^2)\}^{1/2}=O(n^{-1/2}+\zeta_n^2)\]
	if multiple linear regression is used for the estimation of conditional mean and conditional variance. 
\end{thm}

\begin{proof}
	By Theorem \ref{thm:strong} and its proof, using the triangle inequality it suffices to show that
\[[E\{|\hat{m}(z)-m(z)|^2\}]^{1/2}=O(n^{-1/2}+\zeta_n^2),\quad [E\{|\hat{v}^2(z)-v^2(z)|^2\}]^{1/2}=O(n^{-1/2}+\zeta_n^2)\]
for any fixed $z\in\mathbb{R}\times\mathcal{T}$. We will only prove the result for the conditional mean, as the proof for the conditional variance is similar.

	Recall that to estimate the conditional mean function, we use the regression model with two-dimensional predictors $Z_i=(Y_{i1}, T_{i1})^\T$ and responses $Y_i=Y_{i2}$. Viewing the $\{(Z_i, Y_i)\}_{i=1}^n$ as $n$ i.i.d. realizations of the random pair $(Z, Y)$ where $Z=(U, T)^\T$, for multiple linear regression we have
	\[m(Z)=\beta_0+\beta_uU+\beta_tT.\]
	In the presence of measurement errors, we only observe the error-prone predictor $V=U+\zeta_nW$ where $W$ has mean zero and variance 1. The regression model then becomes
	\begin{equation}
		\label{eq:mstar}
		m^*(Z)=\beta_0^*+\beta_u^*V+\beta_t^*T.
	\end{equation}
	In Appendix B.2 of \citet{carr:06}, it is shown that the ordinary least squares estimator of the intercept and coefficients of $V$ and $T$ in \eqref{eq:mstar} is consistent for
	\[\beta_0^*=\beta_0,\quad\beta_u^*=\lambda_1\beta_u,\quad\beta_t^*=\beta_t+(1-\lambda_1)\beta_u\gamma_t,\]
	respectively, with the attenuating factor
	\[\lambda_1=\frac{\sigma_{u|t}^2}{\sigma_{u|t}^2+\zeta_n^2}.\]
	Here $\sigma_{u|t}^2$ is the residual variance of the regression of $U$ on $T$, and $\gamma_t$ is the coefficient of $T$ in the regression of $U$ on $T$, that is, $E(U|T)=\gamma_0+\gamma_tT$.
	
	It follows that
	\begin{align*}
		m^*(z)-m(z)&=(\lambda_1-1)\beta_uU+(1-\lambda_1)\beta_u\gamma_tT\\&=(1-\lambda_1)(\beta_u\gamma_tT-\beta_uU)\\&=\frac{\zeta_n^2}{\sigma_{u|t}^2+\zeta_n^2}(\beta_u\gamma_tT-\beta_uU).
	\end{align*}
	Using the fact that
	\[[E\{|\hat{m}(z)-m^*(z)|^2\}]^{1/2}=O(n^{-1/2})\]
	and 
	\[\hat{m}(z)-m(z)=\hat{m}(z)-m^*(z)+m^*(z)-m(z),\]
	we conclude that
	\[[E\{|\hat{m}(z)-m(z)|^2\}]^{1/2}=O(n^{-1/2}+\zeta_n^2).\]

\end{proof}

%

\section{Additional Simulations}
\label{supp:simu}
\subsection{Simulating snippets from the Berkeley growth study data}
\label{subsec:bgsd}
We illustrate the proposed method by drawing snippet samples from the Berkeley growth study data \citep{tudd:54}, where growth curves quantified by height in cm of 93 children (54 females and 39 males) are measured quarterly from 1 to 2 years, annually from 2 to 8 years and biannually from 8 to 18 years. Again females and males were treated separately due to the sexual dimorphism of growth. To visually evaluate the quality of the estimated sample paths, we generated  synthetic growth curves for females and males, respectively. Specifically, the mean function, the first three eigenfunctions, and functional principal component scores for each female were first estimated from the original growth data for females, where the first three components account for $97.5\%$ of the variation. We then resampled the functional principal component scores from their joint distribution and used the resampled scores to construct an augmented sample of 300 growth curves for females and analogously for males. Artificial snippets were  created by randomly selecting two measurements, one year apart, for each synthetic growth curve.

\single
\begin{figure}[tb]
	\centering
	\includegraphics[width=\linewidth]{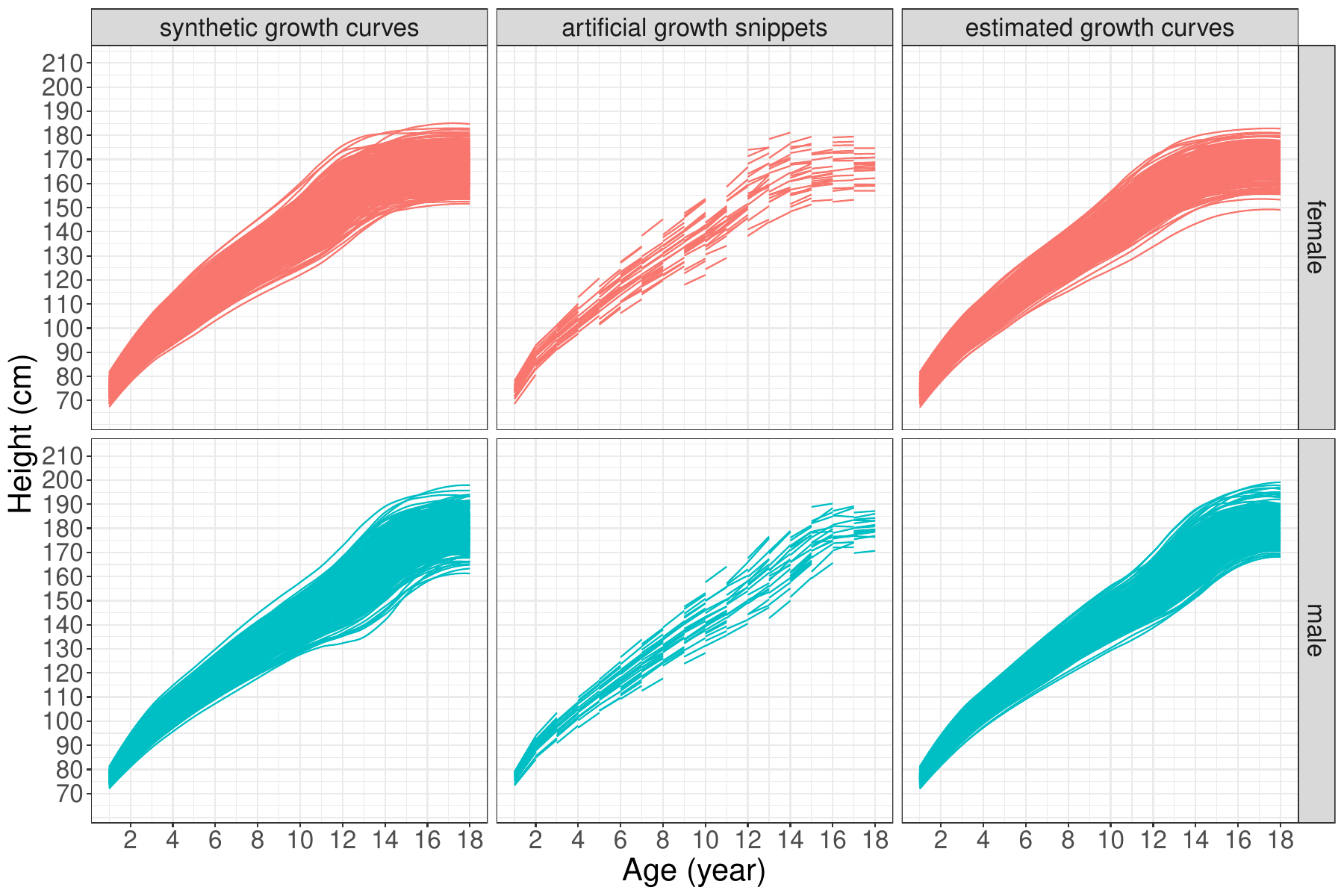}
	\caption{Synthetic growth curves (left), artificial growth snippets (middle), and estimated growth curves (right) for the Berkeley growth study data.}
	\label{fig:bgduncondition}
\end{figure}
\double

To recover the underlying growth dynamics from these snippet samples, we carried out the proposed recursive procedure in \eqref{eq:simest} 1000 times to obtain 1000 estimated growth curves, where local linear regression is utilized to estimate the conditional mean and conditional variance. The starting time is $t_0=1$ year old for both females and males, while the time spacing is the measurement spacing  in the artificial snippets, i.e., $\Delta=1$ year. To reconstruct the unconditional dynamic distributions for the Berkeley growth study data, a starting height $X_0$ was randomly chosen at the age of one year. The bandwidth for both the conditional mean and conditional variance estimation was chosen as $h=(20, 0.5)^\T$ using leave-one-out cross-validation. Figure~\ref{fig:bgduncondition}, from left to right, shows the synthetic growth curves, artificial growth snippets, and estimated growth curves for females and males, respectively. 
We observe that the estimated growth curves recover the underlying growth dynamics from the artificial growth snippets, and the recovered estimates match those for the synthetic growth curves, demonstrating the utility of the proposed method.

\bco
\begin{figure}[tb]
	\centering
	\includegraphics[width=\linewidth]{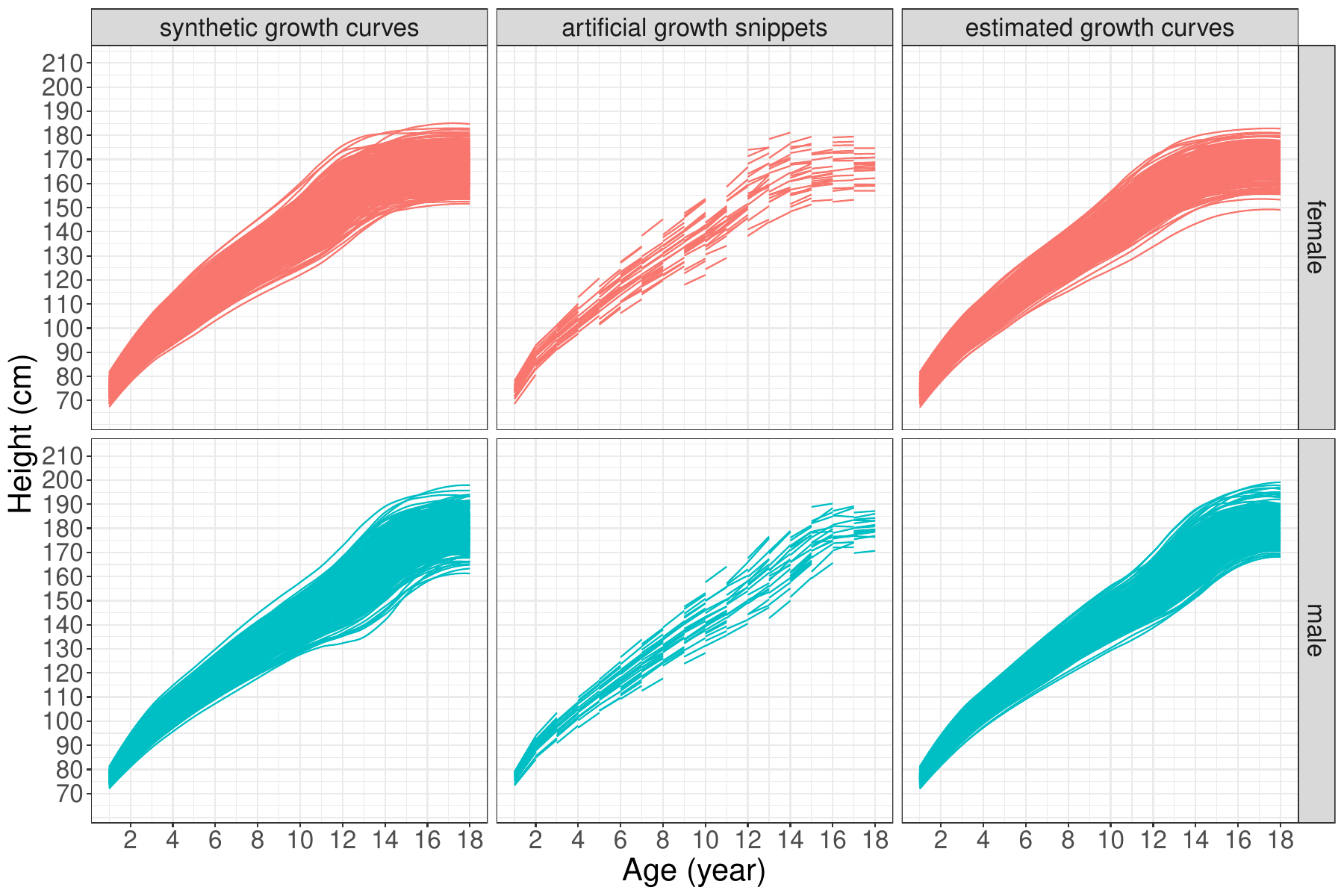}
	\caption{Synthetic growth curves (left), artificial growth snippets (middle), and estimated growth curves (right) for the Berkeley growth study data. The synthetic growth curves with a starting height far away from the average starting height ($\pm0.5$ cm) are colored in steel blue with a lower opacity. The starting time of the estimated growth curves is $t_0=1$ year, with starting heights of 73 cm and 76 cm for females and males, respectively.}
	\label{fig:bgd}
\end{figure}
\fi
\subsection{Simulations for non-Gaussian processes}
The Black-Scholes process, also known as geometric Brownian motion, was introduced by \citet{blac:73} and \citet{mert:73} to model the prices of financial assets. The process is defined as the unique strong solution of 
\[dX_t=\theta X_tdt+\sigma X_tdB_t,\]
with $X_0>0, \sigma>0$. Compared to the Ornstein-Uhlenbeck process, the Black-Scholes process has a more complex diffusion coefficient that changes with the value of the process. In finance, the drift parameter is interpreted as the constant interest rate and the diffusion coefficient as the volatility. 

To assess the robustness of the proposed approach to violations of Gaussianity, we implement the same simulation procedure as in subsection 5.2 for the Black-Scholes process, comparing the proposed approach and the covariance completion approach. The conditional distribution of $X_t$ given $X_s$ is log-normal, leading to the following exact simulation of the Black-Scholes process at $\{t_k\}_{k=0}^K$,
\[X_{k+1}=\mathrm{exp}\{(\theta-\frac{1}{2}\sigma^2)(t_{k+1}-t_k)+\sigma(t_{k+1}-t_k)^{1/2}W_k\}X_k,\]
where the  $W_k\sim N(0, 1)$ are independent for all $k$ and $X_0=x_0$. The mean and standard deviation of RMSE across $Q=500$ runs for various sample sizes and noise levels are summarized in Table~\ref{tab:armsen}. Similar to the Ho-Lee model and Ornstein-Uhlenbeck process in subsection 5.2, the mean RMSE of the proposed approach decreases as sample size increases, suggesting that it is robust to violations of Gaussianity. Conversely, the covariance completion approach does not perform very well, possibly due to the exceptionally sparse nature of the scenario examined.

\single
\begin{table}[tb]
	\footnotesize
	\centering
	\caption{Mean and standard deviation (in parentheses) of root-mean-square errors across 500 runs for the geometric Brownian motion. \textbf{DM}: the proposed dynamic modeling approach; \textbf{LW}: the covariance completion approach by \citet{lin:22}}
	\begin{tabular}{l|c|c|c|c|c|c}
		\hline
		\multirow{2}{*}{\diagbox[height=2\line]{\shortstack{Sample \\ size}}{\shortstack{Noise \\ level \\\vspace{-1em}}}} & \multicolumn{3}{c|}{DM} & \multicolumn{3}{c}{LW}\\\cline{2-7}
		& 0 & 0.01 & 0.1 & 0 & 0.01 & 0.1\\\hline
		50 & 1.13 (2.14) & 1.07 (1.90) & 1.07 (2.30) & 0.78 (0.25) & 0.78 (0.25) & 0.83 (0.36)\\\hline
		200 & 0.40 (0.32) & 0.43 (0.35) & 0.41 (0.24) & 0.66 (0.16) & 0.68 (0.17) & 0.68 (0.16)\\\hline
		1000 & 0.22 (0.12) & 0.22 (0.11) & 0.26 (0.08) & 0.59 (0.17) & 0.59 (0.17) & 0.60 (0.17)\\\hline
	\end{tabular}
	\label{tab:armsen}
\end{table}
\double

\subsection{Simulations for $N_i>2$}
\label{supp:rep}
We also assessed the performance of the proposed approach in scenarios where the number of measurements per subject $N_i$ exceeds 2. Using an Ornstein-Uhlenbeck process, as discussed in subsection 5.2, snippets were generated as $\{(X_{T_i, i}, X_{T_i+\delta, i}, \ldots, X_{T_i+4\delta, i})^\T\}_{i=1}^n$ with noise level $\nu=0$. For each $i$,   $T_i$ was a randomly selected time point from the grid $\{t_k\}_{k=0}^{K-4}$. Despite having $N_i=5$ measurements per subject, the simulated functional data still constitute functional snippets, as each subject is observed over a sub-interval spanning only one-fifth of the domain of interest. 

We employed a similar strategy to that described in subsection 6.1 to fully utilize the dynamic information within functional snippets, where the five measurements are divided into four pairs of contiguous measurements for each $i$. For comparison, we also implemented the proposed approach using only two randomly selected contiguous measurements and the covariance completion approach. The mean and standard deviation of RMSE across $Q=500$ runs for various sample sizes are summarized in Table~\ref{tab:armser}. We observe that the proposed strategy effectively utilizes all available measurements for cases where $N_i>2$ and results in substantially lower mean RMSE compared to the case where only two measurements are utilized. The covariance completion approach, unlike in subsection 5.2, performs reasonably well here due to the increased number of measurements available but is still inferior to the proposed approach.

\single
\begin{table}[tb]
	\centering
	\caption{Mean and standard deviation (in parentheses) of root-mean-square errors across 500 runs for the case $N_i=2$. \textbf{DM}: the proposed dynamic modeling approach using all five measurements; \textbf{DM0}: the proposed dynamic modeling approach using only two randomly selected contiguous measurements; \textbf{LM}: the covariance completion approach by \citet{lin:22} using all five measurements}
	\begin{tabular}{l|c|c|c}
		\hline
		Sample size&DM&DM0&LW\\\hline
		50 & 0.25 (0.14) & 0.61 (0.71) & 0.61 (0.28)\\\hline
		200 & 0.12 (0.06) & 0.25 (0.16) & 0.47 (0.13)\\\hline
		1000 & 0.05 (0.03) & 0.11 (0.05) & 0.41 (0.06)\\\hline
	\end{tabular}
	\label{tab:armser}
\end{table}
\double

\subsection{Simulations for uncertainty quantification}
As discussed in Section 5.1, the proposed dynamic modeling approach inherently provides uncertainty quantification for the estimated sample paths. In this section, we validate the coverage of the empirical confidence band for the Ornstein-Uhlenbeck process.

First, we generate $M=1000$ sample paths using the proposed method and construct the corresponding $95\%$ pointwise confidence band using these $M=1000$ sample paths. We then investigate the coverage of this pointwise confidence band at the last time point $t_K=1$. Specifically, we generate another set of $M=1000$ sample paths and calculate the proportion whose value at $t_K=1$ falls within the previously constructed confidence band. This simulation is repeated $Q=500$ times, with results summarized in Table ~\ref{tab:armseuq}.

We observe that the empirical $95\%$ confidence band constructed using the proposed approach consistently achieves the nominal coverage across all settings considered. This validation demonstrates the method's ability to provide reliable uncertainty quantification for the estimated sample paths.

\single
\begin{table}[tb]
	\centering
	\caption{Mean and standard deviation (in parentheses) of coverage levels across 500 runs. \textbf{DM}: the proposed dynamic modeling approach; \textbf{LW}: the covariance completion approach by \citet{lin:22}}
	\begin{tabular}{l|c|c|c}
		\hline
		\multirow{2}{*}{Sample size} & \multicolumn{3}{c}{Noise level}\\\cline{2-4}
		& 0 & 0.01 & 0.1\\\hline
		50 & 0.948 (0.010) & 0.948 (0.010) & 0.948 (0.010)\\\hline
		200 & 0.949 (0.009) & 0.948 (0.010) & 0.947 (0.010)\\\hline
		1000 & 0.947 (0.010) & 0.948 (0.010) & 0.948 (0.010)\\\hline
	\end{tabular}
	\label{tab:armseuq}
\end{table}
\double

\subsection{Simulations for mean and variance estimation}
While the focus of the proposed approach is to recover the dynamic distribution of the underlying process, we also evaluate its performance in estimating the mean and variance functions using the following $L_2$ errors, 
\begin{align*}
\mathrm{LE}_{\mu}&=\left\{\int_0^1(\hat{\mu}(t)-\mu(t))^2dt\right\}^{1/2},
\\\mathrm{LE}_{\Sigma}&=\left\{\int_0^1(\hat{\Sigma}^{1/2}(t, t)-\Sigma^{1/2}(t, t))^2dt\right\}^{1/2},
\end{align*}
where $\hat{\mu}_q(\cdot)$ and $\hat{\Sigma}_q(\cdot, \cdot)$ denote the sample mean and variance functions of the $M=1000$ estimated sample paths $\{(\hat{X}_{t_1, l}, \ldots, \hat{X}_{t_K, l})^\T\}_{l=1}^M$. The mean and standard deviation (in parentheses) of $L_2$ errors across various sample sizes and noise levels are summarized in Table~\ref{tab:armsem} and Table~\ref{tab:armsev}. We observe that for both Gaussian and non-Gaussian processes, the mean $L_2$ errors decrease with increasing sample sizes, indicating consistency in the estimation of the mean and variance functions.

\single
\begin{table}[tb]
	\centering
	\caption{Mean and standard deviation (in parentheses) of $L_2$ errors for the mean function across 500 runs}
	\begin{tabular}{l|c|c|c}
		\hline
		\multirow{2}{*}{Sample size} & \multicolumn{3}{c}{Noise level}\\\cline{2-4}
		& 0 & 0.01 & 0.1\\\hline
		\multicolumn{4}{l}{\textbf{Ho-Lee model}}\\\hline
		50 & 0.254 (0.633) & 0.273 (0.632) & 0.528 (1.965)\\\hline
		200 & 0.044 (0.067) & 0.046 (0.065) & 0.066 (0.107)\\\hline
		1000 & 0.009 (0.011) & 0.009 (0.012) & 0.012 (0.016)\\\hline
		\multicolumn{4}{l}{\textbf{Ornstein-Uhlenbeck process}}\\\hline
		50 & 0.159 (0.459) & 0.182 (0.634) & 0.209 (1.683)\\\hline
		200 & 0.023 (0.036) & 0.026 (0.036) & 0.025 (0.036)\\\hline
		1000 & 0.004 (0.006) & 0.005 (0.007) & 0.005 (0.006)\\\hline
		\multicolumn{4}{l}{\textbf{Geometric Brownian motion}}\\\hline
		50 & 0.272 (1.686) & 0.260 (1.428) & 0.313 (1.666)\\\hline
		200 & 0.021 (0.036) & 0.024 (0.043) & 0.024 (0.046)\\\hline
		1000 & 0.004 (0.008) & 0.004 (0.006) & 0.004 (0.006)\\\hline
	\end{tabular}
	\label{tab:armsem}
\end{table}
\double

\single
\begin{table}[H]
	\centering
	\caption{Mean and standard deviation (in parentheses) of $L_2$ errors for the variance function across 500 runs}
	\begin{tabular}{l|c|c|c}
		\hline
		\multirow{2}{*}{Sample size} & \multicolumn{3}{c}{Noise level}\\\cline{2-4}
		& 0 & 0.01 & 0.1\\\hline
		\multicolumn{4}{l}{\textbf{Ho-Lee model}}\\\hline
		50 & 0.125 (0.545) & 0.123 (0.401) & 0.322 (2.078)\\\hline
		200 & 0.016 (0.027) & 0.015 (0.024) & 0.059 (0.108)\\\hline
		1000 & 0.003 (0.005) & 0.003 (0.005) & 0.022 (0.023)\\\hline
		\multicolumn{4}{l}{\textbf{Ornstein-Uhlenbeck process}}\\\hline
		50 & 0.064 (0.236) & 0.077 (0.514) & 0.311 (4.953)\\\hline
		200 & 0.008 (0.015) & 0.008 (0.021) & 0.009 (0.017)\\\hline
		1000 & 0.002 (0.002) & 0.002 (0.002) & 0.003 (0.003)\\\hline
		\multicolumn{4}{l}{\textbf{Geometric Brownian motion}}\\\hline
		50 & 0.449 (3.789) & 0.359 (3.082) & 0.579 (5.292)\\\hline
		200 & 0.025 (0.091) & 0.030 (0.086) & 0.031 (0.076)\\\hline
		1000 & 0.005 (0.011) & 0.005 (0.008) & 0.010 (0.012)\\\hline
	\end{tabular}
	\label{tab:armsev}
\end{table}
\double

\section{Spinal bone mineral density data}
\begin{figure}[tb]
	\centering
	\includegraphics[width=\linewidth]{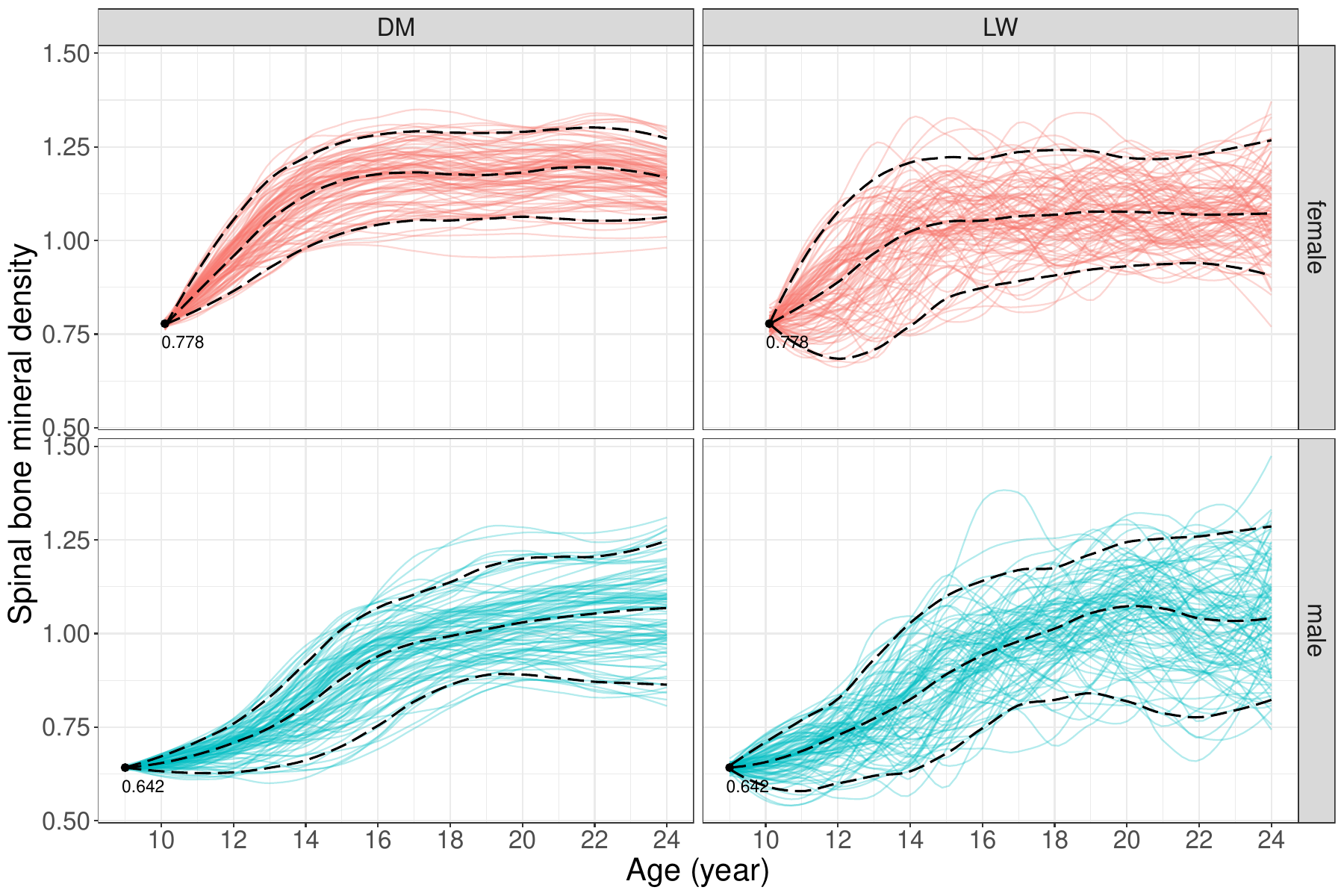}
	\caption{Estimated bone density curves for the spinal bone mineral density data using the proposed dynamic modeling approach (DM) and the covariance completion approach by \citet{lin:22} (LW). The black dashed curves indicate $5\%, 50\%$ and  $95\%$ percentiles. The bone density measurements of the selected female and male are also highlighted.}
	\label{fig:bmdc}
\end{figure}

We applied the covariance completion approach to the spinal bone mineral density data, obtaining 100 estimated bone density curves for the same females and males selected in subsection 6.2. Figure~\ref{fig:bmdc} illustrates these results, alongside those obtained using the proposed approach for comparison. We observe that the covariance completion approach tends to produce excessive variability when recovering the underlying dynamics from functional snippets. The estimated bone density curves using the covariance completion approach exhibit greater variability over time, whereas those produced by the proposed approach are smoother, better reflecting the smooth evolution of the biological process.

\single
\double

\bco
\section{Square-integrable stochastic processes and Karhunen-Lo\`{e}ve Expansion}
\label{supp:kl}
According to Theorem 1, the stochastic process $X_t$, as a solution to SDE (6), is square-integrable. By the Mercer's theorem, the covariance function $\Sigma(s, t)$ admits the following spectral decomposition,
\[\Sigma(s, t)=\sum_{l=1}^\infty\lambda_l\phi_l(s)\phi_l(t),\]
where $\lambda_1\geq\lambda_2\geq\cdots\geq0$ are eigenvalues and $\{\phi_l\}_{l=1}^\infty$ are the corresponding eigenfunctions. The stochastic process $X_t$ admits the so-called Karhunen-Lo\`{e}ve expansion
\[X_t=\mu(t)+\sum_{l=1}^\infty\xi_l\phi_l(t),\]
where $\xi_l=\int_\mathcal{T}\{X_t-\mu(t)\}\phi_l(t)dt$ are uncorrelated zero mean random variables with variance $\lambda_l$.

Under Gaussianity, the drift and diffusion coefficients in (6) are
\begin{align*}
	b(t, X_t)&=\mu'(t)+\Sigma_s'(s, t)\big|_{s=t}\Sigma^{-1}(t, t)\{X_t-\mu(t)\}\\&=\mu'(t)+\frac{\sum_{l=1}^\infty\lambda_l\phi_l(t)\phi'_l(t)}{\sum_{l=1}^\infty\lambda_l\phi_l^2(t)}\{X_t-\mu(t)\},\\
	\sigma(t, X_t)&=\left\{\Sigma'(t, t)-2\Sigma_s'(s, t)\big|_{s=t}\right\}^{1/2}\\&=\{2\sum_{l=1}^\infty\lambda_l\phi_l(t)\phi'_l(t)-2\sum_{l=1}^\infty\lambda_l\phi_l(t)\phi'_l(t)\}^{1/2}.
\end{align*}
\fi

\end{document}